\documentclass[11pt]{amsart}
\usepackage{latexsym,amssymb,amsmath,amscd,amsthm}
\numberwithin{equation}{section}
\theoremstyle{remark}
\newtheorem{theorem}{{\bf THEOREM}}[section]

\newtheorem{corollary}{{\bf COROLLARY}}[section]
\newtheorem{example}{{\bf EXAMPLE}}[section]
\newtheorem{proposition}{{\bf PROPOSITION}}[section]
\newcommand{\bq}{\begin{equation}}
\newcommand{\bea}{\begin{array}}
\newcommand{\eea}{\end{array}}

\newcommand{\ga}{\alpha}
\newcommand{\gep}{\epsilon}
\newcommand{\gD}{\Delta}
\newcommand{\gl}{\lambda}

\newcommand{\gb}{\beta}

\newcommand{\mf}{\mathfrak}
\newcommand{\mc}{\mathcal}

\newcommand{\wg}{\wedge}

\newcommand{\go}{\omega}
\newcommand{\gO}{\Omega}

\newcommand{\gt}{\theta}
\newcommand{\gs}{\sigma}
\newcommand{\gz}{\zeta}
\newcommand{\gag}{\gamma}
\newcommand{\gd}{\delta}
\newcommand{\pp}{\partial}

\newcommand{\tl}{\tilde}
\newcommand{\na}{\nabla}

\newcommand{\bs}{\blacksquare}

\newcommand{{\DDD}}{D\!\!\!\!\!\!-}
\title{REMARKS ON THE SCHR\"ODINGER EQUATION}
\author{Robert Carroll\\University of Illinois, Urbana, IL 61801}

\date{January, 2004\thanks{email: rcarroll@math.uiuc.edu}}

\begin{document}

\bibliographystyle{plain}

\begin{abstract}
Various origins of linear and nonlinear Schr\"odinger equations are discussed in connection
with diffusion, hydrodynamics, and fractal structure.  The treatment is mainly expository,
emphasizing the quantum potential, with a few new observations.
\end{abstract}

\maketitle

\tableofcontents

\section{INTRODUCTION}
\renewcommand{\theequation}{1.\arabic{equation}}
\setcounter{equation}{0}

Perhaps no subject has been the focus of as much mystery as ``classical" quantum mechanics
(QM) even though the standard Hilbert space framework provides an eminently satisfactory
vehicle for determining accurate conclusions in many situations.  This and other classical
viewpoints provide also seven decimal place accuracy in QED for example.  So why all the
fuss?  The erection of the Hilbert space edifice and the subsequent development of operator
algebras (extending now into noncommutative (NC) geometry) has an air of magic.  It works
but exactly why it works and what it really represents remain shrouded in ambiguity.
Also geometrical connections of QM and classical mechanics (CM) are still a source of new
work and a modern paradigm focuses on the emergence of CM from QM (or below).  Below
could mean here a micro structure of space time (quantum foam, Cantorian spacetime, etc.).
In addition there are beautiful stochastic theories for diffusion and QM.  In terms of
background information in book form we mention here e.g. \cite{a1,b89,ch,c1,
c89,c33,c34,c89,h99,k99,l98,l99,m99,m98,m96,m97,n9,n15,n6,n5,p99,r9,s3} (the lecture notes
\cite{c47,c46,c48,c51,c52} in a more polished and organized form should also eventually
become part of a book in preparation).  The present paper focuses on certain aspects of
the Schr\"odinger equation (SE) involving the wave function form $\psi=Rexp(iS/\hbar)$,
hydrodynamical versions, diffusion processes, quantum potentials, and fractal methods.
The aim is to envision ``structure", both mathematical and physical, and we avoid detailed
technical discussion of mathematical fine points (cf. \cite{c33,c34,c36,c56,n10,o3,r9,s6}
for various delicate matters).  Rather than looking at such matters as Markov
processes with jumps for example we prefer to seek ``meaning" for the 
Schr\"odinger equation via
microstructure and fractals in connection with diffusion processes and kinetic theory.

\section{BACKGROUND FOR THE SCHR\"ODINGER EQUATION}
\renewcommand{\theequation}{2.\arabic{equation}}
\setcounter{equation}{0}

First consider the SE in
the form ${\bf (A1)}\,\,-(\hbar^2/2m)\psi''+V\psi=i\hbar\psi_t$ so that for $\psi=
Rexp(iS/\hbar)$ one obtains
\bq\label{2.1}
S_t+\frac{S_X^2}{2m}+V-\frac{\hbar^2R''}{2mR}=0;\,\,\pp_t(R^2)+\frac{1}{m}(R^2S')'=0
\end{equation}
where $S'\sim\pp S/\pp X$.  Writing $P=R^2$ (probability density $\sim |\psi|^2$)
and $Q=-(\hbar^2/2m)(R''/R)$ (quantum potential) this becomes
\bq\label{2.2}
S_t+\frac{(S')^2}{2m}+Q+V=0;\,\,P_t+\frac{1}{m}(PS')'=0
\end{equation}
and this has some hydrodynamical interpretations in the spirit of Madelung.  Indeed  
going to \cite{d3}
for example we take $p=S'$ with $p=m\dot{q}$ for $\dot{q}$ a
velocity (or ``collective" velocity - unspecified).  Then \eqref{2.2} can be written as
($\rho=mP$ is an unspecified mass density)
\bq\label{2.3}
S_t+\frac{p^2}{2m}+Q+V=0;\,\,P_t+\frac{1}{m}(Pp)'=0;\,\,p=S';\,\,P=R^2;\,\,Q=
-\frac{\hbar^2}{2m}\frac{R''}{R}=-\frac{\hbar^2}{2m}\frac{\pp^2\sqrt{\rho}}
{\sqrt{\rho}}
\end{equation}
Note here
\bq\label{2.4}
\frac{\pp^2\sqrt{\rho}}{\sqrt{\rho}}=\frac{1}{4}\left[\frac{2\rho''}{\rho}-
\left(\frac{\rho'}{\rho}\right)^2\right]
\end{equation}
Now from $S'=p=m\dot{q}=mv$ one has
\bq\label{2.5}
P_t+(P\dot{q})'=0\equiv\rho_t+(\rho\dot{q})'=0;\,\,S_t+\frac{p^2}{2m}+V-
\frac{\hbar^2}{2m}\frac{\pp^2\sqrt{\rho}}{\sqrt{\rho}}=0
\end{equation}
Differentiating the second equation in X yields ($\pp\sim\pp/\pp X,\,\,v=\dot{q}$)
\bq\label{2.6}
mv_t+mvv'+\pp V-\frac{\hbar^2}{2m}\pp\left(\frac{\pp\sqrt{\rho}}{\sqrt{\rho}}\right)=0
\end{equation}
Consequently, multiplying by $p=mv$ and $\rho$ respectively in \eqref{2.5} and
\eqref{2.6}, we obtain
\bq\label{2.7}
m\rho v_t+m\rho vv'+\rho\pp
V-\frac{\hbar^2}{2m}\rho\pp\left(\frac{\pp^2\sqrt{\rho}}{\sqrt{\rho}}\right)=0;\,\,
mv\rho_t+mv(\rho'v+\rho v')=0
\end{equation}
Then adding in \eqref{2.7} we get
\bq\label{2.8}
\pp_t(\rho v)+\pp(\rho v^2)+\frac{\rho}{m}\pp V-\frac{\hbar^2}{2m^2}\rho\pp
\left(\frac{\pp^2\sqrt{\rho}}{\sqrt{\rho}}\right)=0
\end{equation}
This is similar to an equation in \cite{d3} (called an ``Euler" equation) and it
definitely has a hydrodynamic flavor (cf. also \cite{g99}).
\\[3mm]\indent
Now go to \cite{p4} and write \eqref{2.6} in the form ($mv=p=S'$)
\bq\label{2.9}
\frac{\pp v}{\pp t}+(v\cdot\na)v=-\frac{1}{m}\na(V+Q);\,\,v_t+vv'=-(1/m)\pp(v+Q)
\end{equation}
The higher dimensional form is not considered here but matters are similar there.
This equation (and \eqref{2.8}) is incomplete as a hydrodynamical equation as a
consequence of a missing term $-\rho^{-1}\na {\mf p}$ where ${\mf p}$ is the pressure
(cf. \cite{l5}).
Hence one ``completes" the equation in the form
\bq\label{2.10}
m\left(\frac{\pp v}{\pp t}+(v\cdot\na)v\right)=-\na(V+Q)-\na F;\,\,mv_t+mvv'=
-\pp(V+Q)-F'
\end{equation}
where ${\bf (A2)}\,\,\na F=(1/R^2)\na {\mf p}$ (or $F'=(1/R^2){\mf p}'$).  By the
derivations above this would then correspond to an extended SE of the form
\bq\label{2.11}
i\hbar\frac{\pp\psi}{\pp t}=-\frac{\hbar^2}{2m}\gD\psi+V\psi+F\psi
\end{equation} 
provided one can determine F in terms of the wave function $\psi$.  One notes that
it a necessary condition here involves $curl grad (F)=0$ or ${\bf (A3)}\,\,
curl(R^{-2}\na {\mf p})=0$ which enables one to take e.g. ${\bf (A4)}\,\,{\mf p}=-b
R^2=-b|\psi|^2$.  For one dimension one writes ${\bf (A5)}\,\,F'=-b(1/R^2)\pp|\psi|^2
=-(2b R'/R)\Rightarrow F=-2blog(R)=-b log(|\psi|^2)$.  Consequently one has a
corresponding SE 
\bq\label{2.12}
i\hbar\frac{\pp\psi}{\pp t}=-\frac{\hbar^2}{2m}\psi''+V\psi-b(log|\psi|^2)\psi
\end{equation}
This equation has a number of nice features discussed in \cite{p4} (but serious 
drawbacks as indicated in \cite{c29} - cf. also \cite{c16,d98,d2,d10,g2,n1,n2,n3}). 
For example
${\bf (A6)}\,\,\psi=\gb G(x-vt)exp(ikx-i\go t)$ is a solution of \eqref{2.12}
with $V=0$ and for $v=\hbar k/m$ one gets ${\bf (A7)}\,\,\psi=cexp[-(B/4)(x-vt+d)^2]
exp(ikx-i\go t)$ where $B=4mb/\hbar^2$.  Normalization $\int_{-\infty}^{\infty}
|\psi|^2=1$ is possible with ${\bf (A8)}\,\,|\psi|^2=\gd_m(\xi)=\sqrt{m\ga/\pi}exp
(-\ga m\xi^2)$ where $\ga=2b/\hbar^2,\,\,d=0,$ and $\xi=x-vt$.  For $m\to\infty$
we see that $\gd_m$ becomes a Dirac delta and this means that motion of a particle with
big mass is strongly localized.  This is impossible for ordinary QM since $exp(ikx
-i\go t)$ cannot be localized as $m\to\infty$.  Such behavior helps to explain the
so-called collapse of the wave function and since superposition does not hold
Schr\"odinger's cat is either dead or alive.  Further $v=k\hbar/m$ is equivalent to the
deBroglie relation $\gl=h/p$ since $\gl=(2\pi/k)=2\pi(\hbar/mv)=2\pi(h/2\pi)(1/p)$.
\\[3mm]\indent
{\bf REMARK 2.1.}
We go now to \cite{k5} and the linear SE in the form ${\bf
(A9)}\,\,i(\pp\psi/\pp t)=-(1/2m)
\gD\psi+U(\vec{r})\psi$; such a situation leads to the Ehrenfest equations which have
the form ${\bf (A10)}\,\,<\vec{v}>=(d/dt)<\vec{r}>$ and $<\vec{r}>=\int
d^3x|\psi(\vec{r},t)|^2\vec{r}$ and ${\bf (A11)}\,\,m(d/dt)<\vec{v}>=\vec{F}(t)$ with
$\vec{F}(t)=-\int d^3x|\psi(\vec{r},t)|^2\vec{\na}U(\vec{r})$.  Thus the quantum
expectation values of position and velocity of a suitable quantum system obey the
classical equations of motion and the amplitude squared is a natural probability
weight.  The result tells us that besides the statistical fluctuations quantum systems
posess an extra source of indeterminacy, regulated in a very definite manner by the
complex wave function.  The Ehrenfest theorem can be extended to many point particle
systems and in \cite{k5} one singles out the kind of nonlinearities that violate the
Ehrenfest theorem.  A theorem is proved that connects Galilean invariance, and the
existence of a Lagrangian whose Euler-Lagrange equation is the SE, to the fulfillment
of the Ehrenfest theorem.$\hfill\bs$
\\[3mm]\indent
{\bf REMARK 2.2.}
There are many problems with the quantum mechanical theory of derived nonlinear SE (NLSE)
but many examples of realistic NLSE arise in the study of superconductivity, Bose-Einstein
condensates, stochastic models of quantum fluids, etc. and the subject demands further
study.  We make no attempt to survey this here but will
give an interesting example later from \cite{c29} related to fractal structures where a
number of the difficulties are resolved.  For further information on NLSE, in addition to
the references above, we refer to \cite{b12,c15,f4,g2,k5,k8,k9,p1,p2,p3,v2,v3} for some
typical situations (the list is not at all complete and we apologize for omissions).
Let us mention a few cases. 
\begin{itemize}
\item
The program of \cite{k5} introduces a Schr\"odinger Lagrangian for a free particle
including self-interactions of any nonlinear nature but no explicit dependence on the
space of time coordinates.  The corresponding action is then invariant under spatial
coordinate transformations and by Noether's theorem there arises a conserved current
and the physical law of conservation of linear momentum.  The Lagrangian is also
required to be a real scalar depending on the phase of the wave function only through
its derivatives.  Phase transformations will then induce the law of conservation of 
probability identified as the modulus squared of the wave function.  Galilean
invariance of the Lagrangian then determines a connection betwee the probability current
and the linear momentum which insures the validity of the Ehrenfest theorem.  
\item
We turn next to \cite{k9} for a statistical origin for QM (cf. also
\cite{ch,c15,k8,k93,n6,o1,r6}).  The idea is to build a program in which the
microscopic motion, underlying QM, is described by a rigorous dynamics different
from Brownian motion (thus avoiding unnecessary assumptions about the Brownian
nature of the underlying dynamics).  The Madelung approach gives rise to fluid
dynamical type equations with a quantum potential, the latter being capable of
interpretation in terms of a stress tensor of a quantum fluid.
Thus one shows in \cite{k9} that the quantum state corresponds to a subquantum
statistical ensemble whose time evolution is governed by classical kinetics in
the phase space.  The equations take the form
\bq\label{2.28}
\rho_t+\pp_x(\rho u)=0;\,\,\pp_t(\mu\rho u_i)+\pp_j(\rho\phi_{ij})+\rho\pp_{x_i}V
=0;\,\,\pp_t(\rho E)+\pp_x(\rho S)-\rho\pp_tV=0
\end{equation}
\bq\label{2.29}
\frac{\pp S}{\pp t}+\frac{1}{2\mu}\left(\frac{\pp S}{\pp x}\right)^2+{\mc W}+
V=0
\end{equation}
for two scalar fields $\rho,\,S$ determining a quantum fluid.  These can be rewritten
as
\bq\label{2.33}
\frac{\pp\xi}{\pp t}+\frac{1}{\mu}\frac{\pp^2S}{\pp x^2}+\frac{1}{\mu}\frac{\pp\xi}
{\pp x}\frac{\pp S}{\pp x}=0;
\end{equation}
$$\frac{\pp S}{\pp t}-\frac{\eta^2}{4\mu}\frac{\pp^2\xi}{\pp
x^2}-\frac{\eta^2}{8\mu}\left(\frac{\pp \xi}{\pp
x}\right)^2+\frac{1}{2\mu}\left(\frac{\pp S}{\pp x}\right)^2+V=0$$
where $\xi=log(\rho)$ and for $\gO=(\xi/2)+(i/\eta)S=log{\Psi}$ with $m=N\mu,\,\, 
{\mc V}=NV$, and $\hbar=N\eta$ one arrives at a SE
\bq\label{2.35}
i\hbar\frac{\pp\Psi}{\pp t}=-\frac{\hbar^2}{2m}\frac{\pp^2 \Psi}{\pp x^2}+{\mc
V}\Psi
\end{equation}
Further one can write $\Psi=\rho^{1/2}exp(i{\mf S}/\hbar)$ with ${\mf S}=NS$ and here
$N=\int|\Psi|^2d^nx$.
The analysis is very
interesting.$\hfill\bs$
\end{itemize}
\indent
{\bf REMARK 2.3.}
Now in \cite{d4} one is obliged to use the form $\psi=Rexp(iS/\hbar)$ to make sense
out of the constructions (this is no problem with suitable provisos, e.g. that S is not
constant - cf. \cite{b3,ch,f2,f3}).  Thus note from
${\bf (A12)}\,\,\psi'/\psi=(R'/R)+i(S'/\hbar)$ with $\Im(\psi'/\psi)=(1/m)S'\sim p/m$
(see also \eqref{3.43} below).  Also note
${\bf (A13)}\,\,J=(\hbar/m)\Im\psi^*\psi'$ and $\rho=R^2=|\psi|^2$ represent a current
and a density respectively.  Then using $p=mv=m\dot{q}$ one can
write ${\bf (A14)}\,\,v=(\hbar/m)\Im(\psi'/\psi)$ and $J=(\hbar/m)\Im|\psi|^2(\psi^*
\psi'/|\psi|^2)=(\hbar/m)\Im(\rho v)$.  Then look at the SE in the form $i\hbar\psi_t=
-(\hbar^2/2m)\psi''+V\psi$ with $\psi_t=(R_t+iS_tR/\hbar)exp(iS/\hbar)$ and 
$\psi_{xx}=[(R'+(iS'R/\hbar)exp(iS/\hbar)]'=[R''+(2iS'R'/\hbar)+(iS''R/\hbar)+
(iS'/\hbar)^2R]exp(iS/\hbar)$ which means
\bq\label{3.41}
-\frac{\hbar^2}{2m}\left[R''-\left(\frac{S'}{\hbar}\right)^2+\frac{2iS'r'}{\hbar}
+\frac{iS''R}{\hbar}\right]+VR=i\hbar\left[R_t+\frac{iS_tR}{\hbar}\right]\Rightarrow
\end{equation}
$$\Rightarrow \pp_tR^2+\frac{1}{m}(R^2S')'=0;\,\,S_t+\frac{(S')^2}{2mR}-
\frac{\hbar^2R''}{2mR}+V=0$$
This can also be written as 
\bq\label{3.42}
\pp_t\rho+\frac{1}{m}\pp(p\rho)=0;\,\,S_t+\frac{p^2}{2m}+Q+V=0
\end{equation}
where $Q=-\hbar^2R''/2mR$.
Now we sketch the philosophy of \cite{d4,d5} in part.  Most of such aspects are omitted
and we try to isolate the essential mathematical features.  First one emphasizes 
configurations based on coordinates whose motion is choreographed by the SE according
to the rule (1-D only here)
\bq\label{3.43}
\dot{q}=v=\frac{\hbar}{m}\Im\frac{\psi^*\psi'}{|\psi|^2}
\end{equation}
where ${\bf (A15)}\,\,i\hbar\psi_t=-(\hbar^2/2m)\psi''+V\psi$.  The argument for 
\eqref{3.43} is based on obtaining the simplest Galilean and time reversal invariant
form for velocity, transforming correctly under velocity boosts.  This leads directly
to \eqref{3.43} ($\sim {\bf (A14)}$) so that Bohmian mechanics (BM) is governed by
\eqref{3.43} and {\bf (A15)}.  It's a fairly convincing argument and no recourse to
Floydian time seems possible (cf. \cite{ch,f3,f7,f8}).  Note however that if $S=c$ then
$\dot{q} =v=(\hbar/m)\Im(R'/R)=0$ while $p=S'=0$ so perhaps this formulation avoids
the $S=0$ problems indicated in \cite{ch,f3,f7,f8}.
One notes also that BM depends only
on the Riemannian structure $g=(g_{ij})=(m_i\gd_{ij})$ in the form ${\bf
(A16)}\,\,\dot{q}=\hbar\Im
(grad\psi/\psi);\,\,i\hbar\psi_t=-(\hbar^2/2)\gD\psi+V\psi$.  What makes the constant
$\hbar/m$ in \eqref{3.43} important here is that with this value the probability
density
$|\psi|^2$ on configuration space is equivariant.  This means that via the evolution
of probability densities $\rho_t+div(v\rho)=0$ (as in \eqref{3.42} with $v\sim p/m$)
the density $\rho=|\psi|^2$ is stationary relative to $\psi$, i.e. $\rho(t)$ retains
the form $|\psi(q,t)|^2$.  One calls $\rho=|\psi|^2$ the quantum equilibrium density
(QED) and says that a system is in quantum equilibrium when its coordinates are
randomly distributed according to the QED.  The quantum equilibrium hypothesis
(QHP) is the assertion that when a system has wave function $\psi$ the distribution
$\rho$ of its coordinates satisfies $\rho=|\psi|^2$.$\hfill\bs$
\\[3mm]\indent
{\bf REMARK 2.4.}
We extract here from \cite{h2,h3,h4} (cf. also the references there for background
and \cite{f1,f10,j1} for some information geometry).
There are a number of interesting results connecting uncertainty, Fisher information,
and QM and we make no attempt to survey the matter.  Thus first recall that the
classical Fisher information associated with translations of a 1-D observable X with
probability density
$P(x)$ is 
\bq\label{5.1}
F_X=\int dx\,P(x)([log(P(x)]')^2>0
\end{equation}
One has a well known Cramer-Rao inequality ${\bf (A17)}\,\,Var(X)\geq F_X^{-1}$
where $Var(X)\sim$ variance of X.  A Fisher length for X is defined via ${\bf
(A18)}\,\,\gd X=F_X^{-1/2}$ and this quantifies the length scale over which $p(x)$
(or better $log(p(x))$) varies appreciably.  Then the root mean square deviation
$\gD X$ satisfies ${\bf (A19)}\,\,\gD X\geq \gd X$.  Let now 
P be the momentum observable conjugate to X, and $P_{cl}$ a classical
momentum observable corresponding to the state $\psi$ given via ${\bf (A20)}\,\,
p_{cl}(x)=(\hbar/2i)[(\psi'/\psi)-(\bar{\psi}'/\bar{\psi})]$ (cf. \eqref{3.43}).   
One has the
identity ${\bf (A21)}\,\,<p>_{\psi}=<p_{cl}>_{\psi}$ following from {\bf (A20)}
with integration by parts. 
Now define the nonclassical momentum by $p_{nc}=p-p_{cl}$ 
and one shows then ${\bf (A21)}\,\,
\gD X\gD p\geq \gd X\gD p\geq \gd X\gD p_{nc}=\hbar/2$.
Now go to \cite{h3} now where two proofs are given for the derivation of the SE from
the exact uncertainty principle (as in {\bf (A21)}).  Thus consider a classical ensemble
of n-dimensional particles of mass m moving under a potential V.  The motion can be
described via the HJ and continuity equations
\bq\label{5.18}
\frac{\pp s}{\pp t}+\frac{1}{2m}|\na s|^2+V=0;\,\,\frac{\pp P}{\pp t}+
\na\cdot\left[P\frac{\na s}{m}\right]=0
\end{equation}
for the momentum potential $s$ and the position probability density P
(note that we have interchanged p and P from \cite{h3} - note also there is no quantum
potential and this will be supplied by the information term).   These
equations follow from the variational principle $\gd L=0$ with Lagrangian
\bq\label{5.19}
L=\int dt\,d^nx\,P\left[\frac{\pp s}{\pp t}+\frac{1}{2m}|\na s|^2+V\right]
\end{equation}
It is now assumed that the classical Lagrangian must be modified due to the existence
of random momentum fluctuations.  The nature of such fluctuations is immaterial for
(cf. \cite{h3} for discussion) and one can assume that the momentum associated with
position x is given by ${\bf (A22)}\,\,p=\na s + N$ where the fluctuation term N
vanishes on average at each point x.  Thus s changes to being an average momentum
potential.  It follows that the average kinetic energy $<|\na s|^2>/2m$ appearing in
\eqref{5.19} should be replaced by $<|\na s+N|^2>/2m$ giving rise to
\bq\label{5.20}
L'=L+(2m)^{-1}\int dt<N\cdot N>=L+(2m)^{-1}\int dt(\gD N)^2
\end{equation}
where $\gD N=<N\cdot N>^{1/2}$ is a measure of the strength of the fluctuations.  The
additional term is specified uniquely, up to a multiplicative constant, by the
following three assumptions
\begin{enumerate}
\item
Action principle:  $L'$ is a scalar Lagrangian with respect to the fields P and s
where the principle $\gd L'=0$ yields causal equations of motion.  Thus ${\bf
(A23)}\,\,(\gD N)^2=\int d^nx\,pf(P,\na P,\pp P/\pp t,s,\na s,\pp s/\pp t,x,t)$ for
some scalar function $f$.
\item
Additivity:  If the system comprises two independent noninteracting subsystems with 
$P=P_1P_2$ then the Lagrangian decomposes into additive subsystem contributions; thus
${\bf (A24)}\,\,f=f_1+f_2$ for $P=P_1P_2$.
\item
Exact uncertainty:  The strength of the momentum fluctuation at any given time is
determined by and scales inversely with the uncertainty in position at that time.  
Thus ${\bf (A25)}\,\,\gD N\to k\gD N$ for $x\to x/k$.  Moreover since position
uncertainty is entirely characterized by the probability density P at any given time
the function $f$ cannot depend on $s$, nor explicitly on $t$, nor on $\pp P/\pp t$.
\end{enumerate}
The following theorem is then asserted (see \cite{c47,h3} for the proofs).
\begin{theorem}
The above 3 assumptions imply ${\bf (A26)}\,\,(\gD N)^2=c\int d^nx\,P|\na
log(P)|^2$ where c is a positive universal constant.
\end{theorem}
\begin{corollary}
It follows from \eqref{5.20} that the equations of motion for $p$ and $s$ corresponding
to the principle $\gd L'=0$ are
\bq\label{5.21}
i\hbar\frac{\pp\psi}{\pp t}=-\frac{\hbar^2}{2m}\na^2\psi+V\psi
\end{equation}
where $\hbar=2\sqrt{c}$ and $\psi=\sqrt{P}exp(is/\hbar)$.$\hfill\bs$
\end{corollary}
\indent
{\bf REMARK 2.5.}
We sketch here for simplicity and clarity another derivation of the SE along similar
ideas following
\cite{r3}.  Let
$P(y^i)$ be a probability density and $P(y^i+\gD y^i)$ be the density resulting from a
small change in the $y^i$.  Calculate the cross entropy via
\bq\label{5.39}
J(P(y^i+\gD y^i):P(y^i))=\int P(y^i+\gD y^i)log\frac{P(y^i+\gD y^i)}{P(y^i)}d^ny\simeq
\end{equation}
$$\simeq\left[\frac{1}{2}\int \frac{1}{P(y^i)}\frac{\pp P(y^i)}{\pp y^i}\frac
{\pp P(y^i)}{\pp y^k)}d^ny\right]\gD y^i\gD y^k=I_{jk}\gD y^i\gD y^k$$
The $I_{jk}$ are the elements of the Fisher information matrix.  The most general
expression has the form
\bq\label{5.40}
I_{jk}(\gt^i)=\frac{1}{2}\int\frac{1}{P(x^i|\gt^i)}\frac{\pp P(x^i|\gt^i)}{\pp \gt^j}
\frac{\pp P(x^i|\gt^i)}{\pp \gt^k}d^nx
\end{equation}
where $P(x^i|\gt^i)$ is a probability distribution depending on parameters $\gt^i$ in
addition to the $x^i$.  For ${\bf (A27)}\,\,P(x^i|\gt^i)=P(x^i+\gt^i)$ one recovers
\eqref{5.39} (straightforward - cf. \cite{r3}).  If P is defined over an n-dimensional
manifold with positive inverse metric $g^{ik}$ one obtains a natural definition of the
information associated with P via
\bq\label{5.41}
I=g^{ik}I_{ik}=\frac{g^{ik}}{2}\int\frac{1}{P}\frac{\pp P}{\pp y^i}\frac{\pp P}{\pp
y^k}d^ny
\end{equation}
Now in the HJ formulation of classical mechanics the equation of motion takes the form
\bq\label{5.42}
\frac{\pp S}{\pp t}+\frac{1}{2}g^{\mu\nu}\frac{\pp S}{\pp x^{\mu}}\frac{\pp S}
{\pp x^{\nu}}+V=0
\end{equation}
where $g^{\mu\nu}=diag(1/m,\cdots,1/m)$.  The velocity field $u^{\mu}$ is given by
${\bf (A28)}\,\,u^{\mu}=g^{\mu\nu}(\pp S/\pp x^{\nu})$.  When the exact coordinates
are unknown one can describe the system by means of a probability density
$P(t,x^{\mu}$ with ${\bf (A29)}\,\,\int Pd^nx=1$ and ${\bf (A30)}\,\,
(\pp P/\pp t)+(\pp/\pp x^{\mu})(Pg^{\mu\nu}(\pp S/\pp x^{\nu})=0$.  These equations
completely describe the motion and can be derived from the Lagrangian
\bq\label{5.43}
L_{CL}=\int P\left\{\frac{\pp S}{\pp t}+\frac{1}{2}g^{\mu\nu}\frac{\pp S}{\pp x^{\mu}}
\frac{\pp S}{\pp x^{\nu}}+V\right\}dtd^nx
\end{equation}
using fixed endpoint variation in S and P.  Quantization is obtained by adding a term
proportional to the information I defined in \eqref{5.41}.  This leads to
\bq\label{5.44}
L_{QM}=L_{CL}+\gl I=\int P\left\{\frac{\pp S}{\pp t}+\frac{1}{2}g^{\mu\nu}\left[
\frac{\pp S}{\pp x^{\mu}}\frac{\pp S}{\pp x^{\nu}}+\frac{\gl}{P^2}\frac{\pp P}
{\pp x^{\mu}}\frac{\pp P}{\pp x^{\nu}}\right]+V\right\}dtd^nx
\end{equation}
Fixed endpoint variation in S leads again to {\bf (A30)} while variation in P leads to
\bq\label{5.45}
\frac{\pp S}{\pp t}+\frac{1}{2}g^{\mu\nu}\left[\frac{\pp S}{\pp x^{\mu}}\frac{\pp S}
{\pp x^{\nu}}+\gl\left(\frac{1}{P^2}\frac{\pp P}{\pp x^{\mu}}\frac{\pp P}{\pp x^{\nu}}
-\frac{2}{P}\frac{\pp^2P}{\pp x^{\mu}\pp x^{\nu}}\right)\right]+V=0
\end{equation}
These equations are equivalent to the Schr\"odinger equation if ${\bf
(A31)}\,\,\psi=\sqrt{P} exp(iS/\hbar)$ with $\gl=(2\hbar)^2$ (cf. Section 6).$\hfill\bs$
\\[3mm]\indent
{\bf REMARK 2.6.}
The SE gives to a probability distribution $\rho=|\psi|^2$ (with suitable
normalization) and to this one can associate an information entropy $S(t)$
(actually configuration information entropy) ${\bf (A32)}\,\,S=-\int\rho log(\rho)
d^3x$ which is typically not a conserved quantity (S is an unfortunate notation
here but we retain it momentarily since no confusion should arise).
The rate of change in time of
S can be readily found by using the continuity equation ${\bf (A33)}\,\,\pp_t\rho=
-\na\cdot(v\rho)$ where $v$ is a current velocity field 
Note here 
(cf. also \cite{p9})
\bq\label{5.49}
\frac{\pp S}{\pp t}=-\int\rho_t(1+log(\rho))dx=\int (1+log(\rho))\pp(v\rho)
\end{equation}
Note that a formal substitution of $v=-u$ in {\bf (A33)} implies the standard free
Browian motion outcome ${\bf (A34)}\,\,dS/dt=D\cdot\int [(\na\rho)^2/\rho)d^3x=
D\cdot Tr{\mf F}\geq 0$ - use ${\bf (A35)}\,\,
u=D\na log(\rho)$ with $D=\hbar/2m$) 
and \eqref{5.49} with $\int
(1+log(\rho))\pp(v\rho)=-\int v\rho\pp log(\rho)=-\int
v\rho'\sim\int((\rho')^2/\rho)$ modulo constants involving D etc.   
Recall here ${mf F}\sim-(2/D^2)\int \rho Qdx=\int dx[(\na \rho)^2/\rho]$ is a functional
form of Fisher information.  A high rate of
information entropy production corresponds to a rapid spreading (flattening down) of
the probablity density.  This delocalization feature is concomitant with the decay in
time property quantifying the time rate at which the far from equilibrium system
approaches its stationary state of equilibrium ${\bf (A36)}\,\,d/dt Tr{\mf F}\leq
0$.$\hfill\bs$ 
\\[3mm]\indent
{\bf REMARK 2.7.}
Now going back to the quantum context one admits general forms of the current
velocity $v$.  For example consider a gradient field $v=b-u$ where the so-called
forward drift $b(x,t)$ of the stochastic process depends on a particular diffusion
model.  Then one can rewrite the continuity equation as a standard Fokker-Plank
equation ${\bf (A37)}\,\,\pp_t\rho=D\gD\rho-\na\cdot(b\rho)$.  Boundary restrictions
requiring $\rho,\,\,v\rho,$ and $b\rho$ to vanish at spatial infinities or at
boundaries yield the general entropy balance equation
\bq\label{5.50}
\frac{dS}{dt}=\int \left[\rho(\na\cdot b)+D\cdot\frac{(\na\rho)^2}{\rho}\right]d^3x
\equiv -D\frac{dS}{dt}=\int \rho(v\cdot u)d^3x=<v\cdot u>
\end{equation}
The first term in the first equation is not positive definite and can be interpreted
as an entropy flux while the second term refers to the entropy production proper. 
The flux term represents the mean value of the drift field divergence $\na\cdot b$
which by itself is a local measure of the flux incoming to or outgoing from
an infinitesimal surrounding of $x$ at time $t$.  If locally $(\na\cdot b)(x,t)>0$ on
an infinitesimal time scale we would encounter a local entropy increase in the system
(increasing disorder) while in case $(\na\cdot b)(x,t)<0$ one thinks of local entropy
loss or restoration or order.  Only in the situation $<\na\cdot b>=0$ is there no
entropy production.
Quantum dynamics permits more complicated behavior.  One looks first for a general
criterion under which the information entropy {\bf (A32)} is a conserved quantity. 
Consider \eqref{2.8} and invoke the diffusion current to write (recall
$u=D(\na\rho)/\rho$)
\bq\label{5.51}
D\frac{dS}{dt}=-\int [\rho^{-1/2}(\rho v)]\cdot[\rho^{-1/2}(D\na\rho)]d^3x
\end{equation}
Then by means of the Schwarz inequality one has ${\bf
(A38)}\,\,D|dS/dt|\leq<v^2>^{1/2}<u^2>^{1/2}$ so a necessary (but insufficient)
condition for $dS/dt\ne 0$ is that both $<v^2>$ and $<u^2>$ are nonvanishing.
On the other hand a sufficient condition for $dS/dt=0$ is that either one of these
terms vanishes.  Indeed in view of ${\bf
(A39)}\,\,<u^2>=D^2\int[(\na\rho)^2/\rho]d^3x$ the vanishing information entropy
production implies $dS/dt=0$; the vanishing diffusion current does the same job.
$\hfill\bs$
\\[3mm]\indent
{\bf REMARK 2.8.}
We develop a little more
perspective now (following \cite{g10} - first paper).  Recall Q 
written out as
\bq\label{5.56}
Q=2D^2\frac{\gD\rho^{1/2}}{\rho^{1/2}}=D^2\left[\frac{\gD
\rho}{\rho}-\frac{1}{2\rho^2}(\na\rho)^2\right]=\frac{1}{2}u^2+D\na\cdot u
\end{equation}
where $u=D\na log(\rho)$ is called an osmotic velocity field.  The standard
Brownian
motion involves $v=-u$, known as the diffusion current velocity and (up to a
dimensional factor) is identified with the thermodynamic force of diffusion which
drives the irreversible process of matter exchange at the macroscopic level.  On
the other hand, even while the thermodynamic force is a concept of purely
statistical origin associated with a collection of particles, in contrast to
microscopic forces which have a direct impact on individual particles themselves,
it is well known that this force manifests itself as a Newtonian type entry in
local conservation laws describing the momentum balance; in fact it pertains to
the average (local average) momentum taken over by the particle cloud, a
statistical ensemble property quantified in terms of the probability distribution
at hand.  It is precisely the (negative) gradient of the above potential Q
in \eqref{5.56} which plays the Newtonian force role in the momentum balance
equations.  The second analytical expression of interest here involves
\bq\label{5.57}
-\int Q\rho dx=(1/2)\int u^2\rho dx=(1/2)D^2\cdot F_X;\,\,F_X=\int\frac{(\na
\rho)^2}{\rho}dx
\end{equation} 
where $F_X$ is the Fisher information, encoded in the probability density $\rho$
which quantifies its gradient content (sharpness plus localization/disorder)
(note $-\int Q\rho=-\int[(1/2)u^2\rho+D\rho u']=-\int (1/2)u^2\rho+\int Du\rho'=
-(1/2)\int D^2(\rho'/\rho)^2\rho+D^2\int \rho'(\rho'/\rho)=(D^2/2)\int (\rho')^2/\rho=
(1/2)\int u^2\rho$).
On the other hand the local entropy production inside the system sustaining an
irreversible process of diffusion is given via 
\bq\label{5.58}
\frac{dS}{dt}=D\cdot\int\frac{(\na\rho)^2}{\rho}dx=D\cdot F_X\geq 0
\end{equation}
This stands for an entropy production rate when the Fick law induced diffusion
current (standard Brownian motion case) $j=-D\na\rho$, obeying $\pp_t\rho+\na
j=0$, enters the scene.  Here $S=-\int \rho log(\rho)dx$ plays the role of (time
dependent) information entropy in the nonequilibrium statistical mechanics
framework for the thermodynamics of irreversible processes.  It is clear that a
high rate of entropy increase coresponds to a rapid spreading (flattening) of the
probability density.  This explicitly depends on the sharpness of density
gradients.  The potential type Q(x,t), the Fisher information $F_X$, the 
nonequilibrium measure of entropy production $dS/dt$, and the information entropy
$S(t)$ are thus mutually entangled quantities, each being exclusively determined
in terms of $\rho$ and its derivatives.
\\[3mm]\indent
In the standard statistical mechanics setting the Euler equation gives a
prototypical momentum balance equation in the (local) mean
\bq\label{5.59}
(\pp_t+v\cdot\na)v=\frac{F}{m}-\frac{\na P}{\rho}
\end{equation}
where $F=-\na F$ represents normal Newtonian force and P is a pressure term.
Q appears in the hydrodynamical formalism of QM via
\bq\label{5.60}
(\pp_t+v\cdot\na)v=\frac{1}{m}F-\na
Q=\frac{1}{m}F+\frac{\hbar^2}{2m^2}\na\frac{\gD\rho^{1/2}}{\rho^{1/2}}
\end{equation}
Another spectacular example pertains to the standard free Brownian motion in
the strong friction regime (Smoluchowski diffusion), namely
\bq\label{5.61}
(\pp_t+v\cdot\na)v=-2D^2\na\frac{\gD\rho^{1/2}}{\rho^{1/2}}=-\na Q
\end{equation}
where $v=-D(\na\rho/\rho)$ (formally $D=\hbar/2m)$.$\hfill\bs$
\\[3mm]\indent
{\bf REMARK 2.9.}
The papers in \cite{d99} contain very interesting derivations of Schr\"odinger equations
via diffusion ideas \`a la Nelson, Markov wave equations, and suitable ``applied" forces
(e.g. radiative reactive forces).$\hfill\bs$

\section{DIFFUSION AND FRACTALS}
\renewcommand{\theequation}{3.\arabic{equation}}
\setcounter{equation}{0}

We go now to Nagasawa \cite{n7,n8,n9,n10} to see how diffusion and the SE are really
connected (cf. also
\cite{a16,b16,c29,g17,n11,n95,n6,o2,o3,o5,o6} for related material, some of which is
discussed later in detail); for now we simply sketch some formulas for a simple
Euclidean metric where
${\bf (B1)}\,\,\gD=\sum(\pp/\pp x^i)^2$. Then 
$\psi(t,x)=exp[R(t,x)+iS(t,x)]$ satisfies a SE
${\bf (B2)}\,\,i\pp_t\psi+(1/2)\gD\psi+ia(t,x)\cdot\na\psi-V(t,x)\psi=0$ 
($\hbar$ and $m$ omitted) if and only if
\bq\label{31}
V=-\frac{\pp S}{\pp t}+\frac{1}{2}\gD R+\frac{1}{2}(\na R)^2-\frac{1}{2}(\na
S)^2-a\cdot\na S;
\end{equation}
$$0=\frac{\pp R}{\pp t}+\frac{1}{2}\gD S+(\na S)\cdot(\na R)+a\cdot\na R$$
in the region ${\bf (B3)}\,\,D=\{(s,x):\,\psi(s,x)\ne 0\}$.
Solutions are often referred to as weak or
distributional but we do not belabor this point.  From \cite{n7,n9} there results
\begin{theorem}
Let $\psi(t,x)=exp[R(t,x)+iS(t,x)]$ be a solution of the SE {\bf (B2)}; then 
${\bf (B4)}\,\,\phi(t,x)=exp[R(t,x)+S(t,x)]$ and $\hat{\phi}=exp[R(t,x)-S(t,x)]$ are
solutions of 
\bq\label{32}
\frac{\pp\phi}{\pp t}+\frac{1}{2}\gD\phi+a(t,x)\cdot\na\phi+c(t,x,\phi)\phi=0;
\end{equation}
$$-\frac{\pp\hat{\phi}}{\pp t}+\frac{1}{2}\gD\hat{\phi}-a(t,x)\cdot\na\hat{\phi}
+c(t,x,\phi)\hat{\phi}=0$$
where the creation and annihilation term $c(t,x,\phi)$ is given via
\bq\label{33}
c(t,x,\phi)=-V(t,x)-2\frac{\pp S}{\pp t}(t,x)-(\na S)^2(t,x)-2a\cdot\na S(t,x)
\end{equation}
Conversely given $(\phi,\hat{\phi})$ as in {\bf (B4)} satisfying \eqref{32}
it follows that $\psi$ satisfies the SE {\bf (B2)} with V as in
\eqref{33} (note $R=(1/2)log(\hat{\phi}\phi)$ and
$S=(1/2)log(\phi/\hat{\phi})$ with $exp(R)=(\hat{\phi}\phi)^{1/2}$).$\hfill\bs$
\end{theorem}
\indent
We will discuss this later in more detail and give proofs 
along with probabilistic content (note that the equations
\eqref{32} are not imaginary time SE).  From this one can conclude that nonrelativistic
QM is diffusion theory in terms of Schr\"odinger processes (described by
$(\phi,\hat{\phi}$) - more details later).  Further it is shown that key postulates in 
Nelson's stochastic mechanics or Zambrini's Euclidean  QM (cf. \cite{z3}) can both be
avoided in connecting the SE to diffusion processes (since they are automatically
valid).  
Look now at Theorem 3.1 for one dimension and write $T=\hbar t$ with $X=(\hbar/\sqrt{m})x$;
then the SE {\bf (B2)} becomes ${\bf
(B5)}\,\,i\hbar\psi_T=-(\hbar^2/2m)\psi_{XX}-iA\psi_X+V\psi$ where $A=a\hbar/\sqrt{m}$.  In
addition ${\bf (B6)}\,\,i\hbar R_T+(\hbar^2/m^2)R_XS_X+(\hbar^2/2m^2)S_{XX}+AR_X=0$ and 
${\bf (B7)} V=-i\hbar S_T+(\hbar^2/2m)R_{XX}+(\hbar^2/2m^2)R_X^2-(\hbar^2/2m^2)S_X^2-AS_X$.
Hence
\begin{proposition}
Equation {\bf (B2)}, written in the variables ${\bf (B8)}\,\,X=(\hbar/\sqrt{m})x,\,\,T=\hbar
t$, with $A=(\sqrt{m}/\hbar)a$ and $V= V(X,T)\sim V(x,t)$ is equivalent to {\bf (B5)}.
\end{proposition}
\indent
Making a change of variables in \eqref{32} now, as in Proposition 3.1, 
yields
\begin{corollary}
Equation \eqref{32}, written in the variables of Proposition 3.1, becomes
\bq\label{37}
\hbar\phi_T+\frac{\hbar^2}{2m}\phi_{XX}+A\phi_X+\tl{c}\phi=0;\,\,-\hbar\hat{\phi}_T+
\frac{\hbar^2}{2m}\hat{\phi}_{XX}-A\hat{\phi}_X+\tl{c}\hat{\phi}=0;
\end{equation}
$$\tl{c}=-\tl{V}(X,T)-2\hbar S_T-\frac{\hbar^2}{m}S_X^2-2AS_X$$
Thus the diffusion processes pick up factors of $\hbar$ and $\hbar/\sqrt{m}$.$\hfill\bs$
\end{corollary}
\indent
{\bf REMARK 3.1.}
We extract here from the Appendix to \cite{n9} for some remarks on competing points of
view regarding diffusion and the the SE.
First some work of Fenyes \cite{f14} is
cited where a Lagrangian is taken as
\bq\label{38}
L(t)=\int\left[\frac{\pp S}{\pp t}+\frac{1}{2}(\na S)^2+V+\frac{1}{2}\left(\frac{1}{2}
\frac{\na\mu}{\mu}\right)^2\right]\mu dx
\end{equation}
where $\mu_t(x)=exp(2R(t,x))$ denotes the distribution density of a diffusion process
and V is a potential function.  The term ${\bf
(B9)}\,\,\Pi(\mu)=(1/2)[(1/2)(\na\mu/\mu)]^2$ is called a diffusion pressure and since
$(1/2)(\na\mu/\mu)\sim\na R$ the Lagrangian can be written as 
\bq\label{39}
L=\int\left[\frac{\pp S}{\pp t}+\frac{1}{2}(\na S)^2+\frac{1}{2}(\na R)^2+V\right]\mu dx
\end{equation}
Applying the variational principle $\gd\int_a^bL(t)dt=0$ one arrives at
\bq\label{310}
\frac{\pp S}{\pp t}+\frac{1}{2}\left[(\na (R+S)\right]^2-(\na(R+S))\cdot\left(
\frac{1}{2}\frac{\na\mu}{\mu}\right)+\left(\frac{1}{2}\frac{\na\mu}{\mu}\right)^2-
\frac{1}{4}\frac{\gD\mu}{\mu}+V=0
\end{equation}
which is called a motion equation of probability densities.  From this he shows that
the function $\psi=exp(R+iS)$ satisfies the SE ${\bf
(B10)}\,\,i\pp_t+(1/2)\gD\psi-V(t,x)\psi=0$.  Indeed putting {\bf (B9)} and the formula
${\bf (B11)}\,\,(1/2)(\gD\mu/\mu)+(1/2)\gD R+(\na R)^2$ into \eqref{39} one obtains
\bq\label{311}
\frac{\pp S}{\pp t}+\frac{1}{2}(\na S)^2-\frac{1}{2}(\na R)^2-\frac{1}{2}\gD R+V=0
\end{equation}
which goes along with the duality relation
${\bf (B12)}\,\,R_t+(1/2)\gD S+\na S\cdot\na R+b\cdot\na R=0$ where ${\bf (B13)}\,\,
u=(1/2)(a+\hat{a})=\na R$ and $v=(1/2)(a-\hat{a})=\na S$ as derived in the Nagasawa
theory. 
Hence $\psi=exp(R+iS)$
satisfies the SE by previous calculations.  One can see however that the equation
\eqref{39} is not needed since the SE and diffusion equations are equivalent and
in fact the equations of motion are the diffusion equations.  Moreover it is shown in
\cite{n9} that \eqref{39} is an automatic consequence in diffusion theory with 
$V=-c-2S_t-(\na S)^2$ and therefore it need not be postulated or derived by other means.
This is a simple calculation from the theory developed above.$\hfill\bs$
\\[3mm]\indent
{\bf REMARK 3.2.}
Nelson's important work in stochastic mechanics \cite{n6} produced the SE from diffusion
theory but involved a stochastic Newtonian equation which is shown in \cite{n9} to be
automatically true.  Thus Nelson worked in a general context which for our purposes here
can be considered in the context of Brownian motions 
\bq\label{312}
B(t)=\pp_t+(1/2)\gD+b\cdot\na+a\cdot\na;\,\,\hat{B}(t)=-\pp_t+(1/2)\gD-b\cdot\na+
\hat{a}\cdot\na
\end{equation}
and used a mean acceleration ${\bf
(B14)}\,\,\ga(t,x)=-(1/2)[B(t)\hat{B}(t)x+\hat{B}(t)B(t)x]$.  Assuming the duality
relations {\bf (B12)} - {\bf (B13)} he obtains a formula
\bq\label{313}
\ga(t,x)=-\frac{1}{2}[B(t)(-b+\hat{a})+\hat{B}(b+a)]=b_t+(1/2)\na(b)^2-(b+v)\times
curl(b)-
\end{equation}
$$-[-v_t+(1/2)\gD u+(1/2)(\hat{a}\cdot\na)a+(1/2)(a\cdot\na)\hat{a}-(b\cdot\na)v-
(v\cdot\na)b-v\times curl(b)]$$
Then it is shown that the SE can be deduced from the stochastic Newton's equation
\bq\label{314}
\ga(t,x)=-\na V+\frac{\pp b}{\pp t}+\frac{1}{2}\na(b^2)-(b+v)\times curl(b)
\end{equation}
Nagasawa shows that this serves only to reproduce a known formula for V
yielding the SE; he also shows that \eqref{313} also is an automatic consequence of
the duality formulation of diffusion equations above.  This equation \eqref{313}  is
often called stochastic quantization since it leads to the SE and it is in fact correct
with the V specified there.  However the SE is more properly considered as following
directly from the diffusion equations in duality and is not correctly an equation of
motion.
There is another discussion of Euclidean QM developed by Zambrini \cite{z3}.  This
involves ${\bf (B15)}\,\,\tl{\ga}(t,x)=(1/2)[B(t)B(t)x+\hat{B}(t)\hat{B}(t)x]$ (with
$(\gs\gs^T)^{ij}=\gd^{ij}$).  It is postulated that this equals ${\bf (B16)}\,\,-\na
c+b_t+(1/2)
\na(b)^2-b+v)\times curl(b)$ which in fact leads to the same equation for V as above
with
$V=-c-2S_t- (\na S)^2-2b\cdot\na S$ so there is nothing new.  Indeed it is shown in
\cite{n9} that {\bf (B16)} holds automatically as a simple consequence of time reversal
of diffusion processes.$\hfill\bs$

\subsection{SCALE RELATIVITY}

There are several excellent and exciting approaches here.  The method of Nottale
\cite{n4,n5,n13} is preeminent (cf. also \cite{o2,o5,o6,o7}) and there is also a nice
derivation of a nonlinear SE via fractal considerations in \cite{c29} (indicated
below).  The most elaborate and rigorous approach is due to Cresson \cite{c13}, with
elaboration and updating in \cite{a10, c31,c32}.
We refer here to \cite{c46,c47,c20,c13,c31,n4,n5}.  There are various derivations of the
SE and we follow \cite{n5} here (cf. also \cite{n13,s99}).  The philosophy is
discussed in \cite{c47,c46,c13,c31,n5} and we just write down equations here.  First a
bivelocity structure is defined (recall that one is dealing with fractal
paths).  One defines first 
\bq\label{315}
\frac{d_{+}}{dt}y(t)=lim_{\gD t\to 0_{+}}\left<\frac{y(t+\gD t)-y(t)}{\gD
t}\right>;
\end{equation}
$$\frac{d_{-}}{dt}y(t)=lim_{\gD t\to 0_{+}}\left<\frac{y(t)-y(t-\gD t)}{\gD
t}\right>$$
Applied to the position vector x this yields forward and backward mean velocities,
namely
${\bf (B17)}\,\,(d_{+}/dt)x(t)=b_{+}$ and $(d_{-}/dt)x(t)=b_{-}$.  Here these
velocities are defined as the average at a point q and time t of the respective
velocities of the outgoing and incoming fractal trajectories; in stochastic QM
this corresponds to an average on the quantum state.  The position vector $x(t)$
is thus ``assimilated" to a stochastic process which satisfies respectively
after ($dt>0$) and before ($dt<0$) the instant t a relation ${\bf (B18)}\,\,
dx(t)=b_{+}[x(t)]dt+d\xi_{+}(t)=b_{-}[x(t)]dt+d\xi_{-}(t)$ where $\xi(t)$ is a
Wiener process (cf. \cite{n6}).  It is in the description of $\xi$ that the $D=2$
fractal character of trajectories is inserted; indeed that $\xi$ is a Wiener
process means that the $d\xi$'s are assumed to be Gaussian with mean 0, mutually
independent, and such that
\bq\label{316}
<d\xi_{+i}(t)d\xi_{+j}(t)>=2{\mc
D}\gd_{ij}dt;\,\,<d\xi_{-i}(t)d\xi_{-j}(t)>=-2{\mc D}\gd_{ij}dt
\end{equation}
where $<\,\,>$ denotes averaging and ${\mc D}$ is the diffusion coefficient.
Nelson's postulate (cf. \cite{n6}) is that ${\mc D}=\hbar/2m$ and this has
considerable justification (cf. \cite{n5}).  Note also that \eqref{316} is indeed
a consequence of fractal (Hausdorff) dimension 2 of trajectories follows from
$<d\xi^2>/dt^2=dt^{-1}$, i.e. precisely Feynman's result $<v^2>^{1/2}\sim \gd
t^{-1/2}$ (the discussion here in \cite{n5} is unclear however - cf. \cite{a11}).  Note
also that  Brownian motion (used in Nelson's postulate) is known to be of fractal
(Hausdorff)
dimension 2. Note also that any value of ${\mc D}$ may lead to QM and for ${\mc D}\to 0$
the theory becomes equivalent to the Bohm theory.
Now expand any function $f(x,t)$ in a Taylor series up to order 2, take averages,
and use properties of the Wiener process $\xi$ to get
\bq\label{317}
\frac{d_{+}f}{dt}=(\pp_t+b_{+}\cdot\na+{\mc D}\gD)f;\,\,\frac{d_{-}f}{dt}=(\pp_t
+b_{-}\cdot\na-{\mc D}\gD)f
\end{equation}
Let $\rho(x,t)$ be the probability density of $x(t)$; it is known that for any
Markov (hence Wiener) process one has ${\bf (B19)}\,\,\pp_t\rho+div(\rho
b_{+})={\mc D}\gD\rho$ (forward equation) and ${\bf (B20)}\,\,\pp_t\rho+div(\rho
b_{-})=-{\mc D}\gD\rho$ (backward equation).  These are called Fokker-Planck
equations and one defines two new average velocities ${\bf (B21)}\,\,
V=(1/2)[b_{+}+b_{-}]$ and $U=(1/2)[b_{+}-b_{-}]$.  Consequently adding and
subtracting one obtains ${\bf (B22)}\,\,\rho_t+div(\rho V)=0$ (continuity equation)
and ${\bf (B23)}\,\,div(\rho U)-{\mc D}\gD\rho=0$ which is equivalent to
${\bf (B24)}\,\,div[\rho(U-{\mc D}\na log(\rho))]=0$.
One can show, using \eqref{317} that the term in square brackets in {\bf (B24)}
is zero leading to ${\bf (B25)}\,\,U={\mc D}\na log(\rho)$.  Now place oneself in
the $(U,V)$ plane and write ${\bf (B26)}\,\,{\mc V}=V-iU$.  Then write
${\bf (B27)}\,\,(d_{{\mc V}}/dt)=(1/2)(d_{+}+d_{-})/dt$ and $(d_{{\mc U}}/dt)=
(1/2)(d_{+}-d_{-})/dt$.    
Combining the equations in \eqref{317} one defines
${\bf (B28)}\,\,(d_{{\mc V}}/dt)=\pp_t+V\cdot\na$ and $(d_{{\mc U}}/dt)={\mc
D}\gD+U\cdot \na$; then define
a complex operator ${\bf (B29)}\,\,(d'/dt)=(d_{{\mc V}}/dt)-i(d_{{\mc U}}/dt)$
which becomes 
\bq\label{318}
\frac{d'}{dt}=\left(\frac{\pp}{\pp t}-i{\mc D}\gD\right)+{\mc V}\cdot\na
\end{equation}
\indent
One now postulates that the passage from classical mechanics to a new
nondifferentiable process considered here can be implemented by the unique
prescription of replacing the standard $d/dt$ by $d'/dt$.  Thus
consider ${\bf (B30)}\,\,{\mc S}=\left<\int_{t_1}^{t_2}{\mc L}(x,{\mc
V},t)dt\right>$ yielding by least action ${\bf (B31)}\,\,(d'/dt)(\pp{\mc
L}/\pp{\mc V}_i)=
\pp{\mc L}/\pp x_i$.  Define then ${\mc P}_i=\pp{\mc L}/\pp{\mc V}_i$ leading to
${\bf (B32)}\,\,{\mc P}=\na {\mc S}$ (note this is ${\mc S}$ and not S).  
Now for Newtonian mechanics write ${\bf
(B33)}\,\, L(x,v,t)=(1/2)mv^2-{\bf U}$ which becomes ${\mc L}(x,{\mc
V},t)=(1/2)m{\mc V}^2-{\mf U}$ leading to ${\bf (B34)}\,\,-\na{\mf U}=m(d'/dt){\mc
V}$. One separates real and imaginary parts of the complex acceleration
$\gag=(d'{\mc V}/dt$ to get
\bq\label{319}
d'{\mc V}=(d_{{\mc V}}-id_{{\mc U}})(V-iU)=(d_{{\mc V}}V-d_{{\mc U}}U)-
i(d_{{\mc U}}V+d_{{\mc V}}U)
\end{equation}
The force $F=-\na{\mf U}$ is real so the imaginary part of the complex
acceleration vanishes; hence
\bq\label{320}
\frac{d_{{\mc U}}}{dt}V+\frac{d_{{\mc V}}}{dt}U=\frac{\pp U}{\pp t}+U\cdot\na
V+V\cdot\na U+{\mc D}\gD V=0
\end{equation}
from which $\pp U/\pp t$ may be obtained.  Differentiating the expression $U={\mc
D}\na log(\rho)$ and using the continuity equation yields another expression
${\bf (B35)}\,\,(\pp U/\pp t)=-{\mc D}\na(div V)-\na(V\cdot U)$.  Comparison
of these relations yields $\na (div V)=\gD V-U\wg curl V$ where the $curl U$ term
vanishes since U is a gradient.  However in the Newtonian case ${\mc P}=m{\mc V}$
so {\bf (B32)} implies that ${\mc V}$ is a gradient and hence a generalization of
the classical action S can be defined via ${\bf (B36)}\,\,V=2{\mc D}\na S$
(note then $\na(div V)=\gD V$ and $curl V=0$). 
Combining this with the expression for U one obtains ${\bf (B37)}\,\,{\mc S}=
log(\rho^{1/2})+iS$.  One notes that this is compatible with \cite{n6} for
example.  The way to the SE is now short; set ${\bf
(B38)}\,\,\psi=\sqrt{\rho}exp(iS)=exp(i{\mc S})$ with
${\bf (B39)}\,\,{\mc V}=-2i{\mc D}\na(log\psi)$
(note $U={\mc D}\na log(\rho),\,\,V=2{\mc D}\na S,\,\,{\mc V}=-2i{\mc
D}\na log{\psi}=-i{\mc D}\na log(\rho)+2{\mc D}\na S=V-iU$); thus for
${\mc P}=m{\mc V}$ the relation ${\bf (B40)}\,\,{\mc P}\sim-i\hbar\na$ or ${\mc
P}\psi=-i\hbar\na\psi$ has a natural interpretation.  Putting $\psi$ in {\bf (B34)},
which generalizes Newton's law to fractal space the equation of motion takes the form
${\bf (B41)}\,\,
\na{\mf U}=2i{\mc D}m(d'/dt)(\na log(\psi))$.  Noting that $d'$ and $\na$ do not
commute one replaces $d'/dt$ by \eqref{318} to obtain
\bq\label{321}
\na{\mf U}=2i{\mc D}m\left[\pp_t\na log(\psi)-i{\mc D}\gD(\na log(\psi))-2i{\mc
D}(\na log(\psi)\cdot\na)(\na log(\psi)\right]
\end{equation}
This expression can be simplified via
\bq\label{322}
\na\gD=\gD\na;\,\,(\na f\cdot\na)(\na f)=(1/2)\na(\na f)^2;\,\,\frac{\gD f}{f}=\gD
log(f)+(\na log(f))^2
\end{equation}
This implies
\bq\label{323}
\frac{1}{2}\gD(\na log(\psi))+(\na log(\psi)\cdot\na)(\na
log(\psi))=\frac{1}{2}\na\frac{\gD\psi}{\psi}
\end{equation}
Integrating this equation yields ${\bf (B42)}\,\,{\mc D}^2\gD\psi+i{\mc
D}\pp_t\psi-({\mf U}/2m)\psi=0$ up to an arbitrary phase factor $\ga(t)$ which can
be set equal to 0 by a suitable choice of phase S.  Replacing ${\mc D}$ by
$\hbar/2m$ one arrives at the SE ${\bf (B43)}\,\,i\hbar\psi_t=-(\hbar^2/2m)\gD\psi
+{\mf U}\psi$.  This suggests an interpretation of QM as mechanics in a
nondifferentiable (fractal) space. 
\\[3mm]\indent
{\bf REMARK 3.3.}
Some of the relevant equations for dimension one are collected together in Section 6.
We note that it is the presence of $\pm$ derivatives that makes possible the introduction
of a complex plane to describe velocities and hence QM; one can think of this as the
motivation for a complex valued wave function and the nature of the SE.  
$\hfill\bs$
\\[3mm]\indent
We go now to \cite{c29} and will sketch some of the material.  Here one extends ideas
of Nottale and Ord in order to derive a nonlinear Schr\"odinger equation (NLSE).
Using the hydrodynamic model
in \cite{p4} one added a hydrostatic pressure term to the Euler-Lagrange equations
and another possibility is to add instead a kinematic pressure term.  The hydrostatic
pressure is based on an Euler equation $-\na p=\rho g$ where $\rho$ is density and 
$g$ the gravitational acceleration (note this gives $p=\rho gx$ in 1-D).  In \cite{p4}
one took $\rho=\psi^*\psi$, b a mass-energy parameter, and $p=\rho$; then the
hydrostatic potential is (for $\rho_0=1$)
\bq\label{330}
b\int g(x)\cdot dr=-b\int\frac{\na p}{\rho}\cdot
dr=-blog(\rho/\rho_0)=-blog(\psi^*\psi)
\end{equation}
Here $-blog(\psi^*\psi)$ has energy units and explains the nonlinear term of
\cite{b99} which involved
\bq\label{331}
i\hbar\frac{\pp\psi}{\pp t}=-\frac{\hbar^2}{2m}\na^2\psi+U\psi-b[log(\psi^*\psi)]\psi
\end{equation}
A derivation of this equation from the Nelson stochastic QM was given by Lemos (cf.
\cite{l99}).  There are however some problems since this equation does not obey the
homogeneity condition saying that the state $\gl|\psi>$ is equivalent to $|\psi>$;
however \eqref{331} is not invariant under $\psi\to\gl\psi$.  Further, plane wave
solutions to \eqref{331} do not seem to have a physical interpretion due to extraneous
dispersion relations.  Finally one would like to have a SE in terms of $\psi$ alone.
Note that another NLSE could be obtained by adding kinetic pressure terms $(1/2)\rho
v^2$ and taking $\rho=a\psi^*\psi$ where $v=p/m$.
Now using the relations from HJ theory ${\bf (B44)}\,\,(\psi/\psi^*)=exp[2i{\mf S}(x)/
\hbar]$ and $p=\na{\mf S}(x)=mv$ one can write ${\bf (B45)}\,\,v=-i(\hbar/2m)\na log
(\psi/\psi^*)$ so that the energy density becomes ${\bf (B46)}\,\,(1/2)\rho|v|^2=
(a\hbar^2/8m^2)\psi\psi^*\na log(\psi/\psi^*)\cdot\na log(\psi^*/\psi)$.  This leads to
a corresponding nonlinear potential associated with the kinematical pressure via
${\bf (B47)}\,\,(a\hbar^2/8m^2)\na log(\psi/\psi^*)\cdot\na log(\psi^*/\psi)$.  Hence a
candidate NLSE is
\bq\label{332}
i\hbar\pp_t=-\frac{\hbar^2}{2m}\na^2\psi+U\psi-b[log(\psi^*\psi)]\psi+
\frac{a\hbar^2}{8m^2}\left(\na log\frac{\psi}{\psi^*}\cdot\na
log\frac{\psi^*}{\psi}\right)
\end{equation}
(apparently this equation has not yet been derived in the literature).
Here the Hamiltonian is Hermitian and $a\ne b$ are both mass-energy parameters to be
determined experimentally.  The new term can also be written in the form ${\bf
(B48)}\,\,\na log(\psi/\psi^*)\cdot\na log(\psi^*/\psi)=-[\na log(\psi/\psi^*)]^2$.
The goal now is to derive a NLSE directly from fractal space time dynamics for a
particle undergoing Brownian motion.  This does not require
a quantum potential,
a hydrodynamic model, or any pressure terms as above.
\\[3mm]\indent
\indent
{\bf REMARK 3.4.}
One should make some comments about the kinematic pressure terms ${\bf
(B49)}\,\,(1/2)\rho v^2\iff (\hbar^2/2m)(a/m)|\na log(\psi)|^2$ versus hydrostatic
pressure terms of the form ${\bf (B50)}\,\,\int (\na p/\rho)\iff -blog(\psi^*\psi)$. 
The hydrostatic term breaks homogeneity whereas the kinematic pressure term
preserves homogeneity (scaling with a $\gl$ factor).  The hydrostatic pressure term
is also not compatible with the motion kinematics of a particle executing a fractal
Brownian motion.  The fractal formulation will enable one to relate the parameters 
$a,b$ to $\hbar$.$\hfill\bs$
\\[3mm]\indent
Following Nottale nondifferentiability implies a loss of causality
and one is thinking of Feynmann paths with $<v^2>\propto (dx/dt)^2\propto
dt^{2[(1/D)-1)}$ with $D=2$.  Now a fractal function $f(x,\gep)$ could have a
derivative $\pp f/\pp\gep$ and renormalization group arguments lead to ${\bf (B51)}\,\,
(\pp f(x,\gep)/\pp log\gep)=a(x)+bf(x,\gep)$ (cf. \cite{n5}).  This can be integrated
to give ${\bf (B52)}\,\,f(x,\gep)=f_0(x)[1-\gz(x)(\gl/\gep)^{-b}]$.
Here $\gl^{-b}\gz(x)$ is an integration constant and $f_0(x)=-a(x)/b$.  This says that
any fractal function can be approximated by the sum of two terms, one independent of
the resolution and the other resolution dependent; one expects $\gz(x)$ to be a
flucuating function with zero mean.  Provided $a\ne 0$ and $b<0$ one has two
interesting cases (i) $\gep<<\gl$ with $f(x,\gep)\sim f_0(x)(\gl/\gep)^{-b}$ and 
(ii) $\gep>>\gl$ with f independent of scale.  Here $\gl$ is the deBroglie
wavelength. Now one writes
\bq\label{333}
r(t+dt,dt)-r(t,dt)=b_{+}(r,t)dt+\xi_{+}(t,dt)\left(\frac{dt}{\tau_0}\right)^{\gb};
\end{equation}
$$r(t,dt)-r(t-dt,dt)-b_{-}(r,t)dt+\xi_{-}(t,dt)\left(\frac{dt}{\tau_0}\right)^{\gb}$$
where $\gb=1/D$ and $b_{\pm}$ are average forward and backward velocities.  This leads
to ${\bf (B53)}\,\,v_{\pm}(r,t,dt)=b_{\pm}(r,t)+\xi_{\pm}(t,dt)(dt/\tau_0)^{\gb-1}$.
In the quantum case $D=2$ one has $\gb=1/2$ so $dt^{\gb-1}$ is a divergent quantity
(so nondifferentiability ensues).  Following \cite{l99,n5,n6} one defines
\bq\label{334}
\frac{d_{\pm}r(t)}{dt}=lim_{\gD t\to\pm 0}\left<\frac{r(t+\gD t)-r(t)}{\gD t}\right>
\end{equation}
from which ${\bf (B54)}\,\,d_{\pm}r(t)/dt=b_{\pm}$.  Now following Nottale one writes
\bq\label{335}
\frac{\gd}{dt}=\frac{1}{2}\left(\frac{d_{+}}{dt}+\frac{d_{-}}{dt}\right)-\frac{i}{2}
\left(\frac{d_{+}}{dt}-\frac{d_{-}}{dt}\right)
\end{equation}
which leads to ${\bf (B55)}\,\,(\gd/dt)=(\pp/\pp t)+v\cdot\na-iD\na^2$.
Here in principle D is a real valued diffusion constant to be related to $\hbar$.
(A symbol D for the fractal dimension is no longer needed here (?) - e.g. $D=2$ with
${\bf (B56)}\,\,<d\xi_{\pm i}d\xi_ {\pm j}>=\pm 2D\gd_{ij}dt$.)  Now for the
complex time dependent wave function we take
$\psi=exp[i{\mf S}/2mD]$ with $p=\na {\mf S}$ so that ${\bf (B57)}\,\,v=-2iD\na log
(\psi)$.  The SE is obtained from the Newton equation ($F=ma$) via ${\bf (B58)}\,\,
-\na U=m(\gd/dt)v=-2imD(\gd/dt)\na log(\psi)$.  Inserting {\bf (B55)} gives
\bq\label{336}
-\na U=-2im[D\pp_t\na log(\psi)]-2D\na\left(D\frac{\na^2\psi}{\psi}\right)
\end{equation}
(see \cite{n5} for identities involving $\na$).  Integrating \eqref{336}
yields ${\bf (B59)}\,\,D^2\na^2\psi+iD\pp_t\psi-(U/2m)\psi=0$ up to an arbitrary phase
factor which may be set equal to zero.  Now replacing D by $\hbar/2m$ one gets the 
SE ${\bf (B60)}\,\,i\hbar\pp_t\psi+(\hbar^2/2m)\na^2\psi=U\psi$.  Here the Hamiltonian
is Hermitian, the equation is linear, and the equation is homogeneous of degree 1 under
the substitution $\psi\to\gl\psi$.
\\[3mm]\indent
Next one generalizes this by relaxing the assumption that the diffusion coefficient
is real.  Some comments on complex energies are needed - in particular constraints
are often needed (cf. \cite{p1}).  However complex energies are not alien in ordinary
QM (cf. \cite{c29} for references).  Now the imaginary part of the linear SE yields
the continuity equation $\pp_t\rho+\na\cdot(\rho v)=0$ and with a complex potential
the imaginary part of the potential will act as a source term in the continuity
equation. Instead of ${\bf (B61)}\,\,<d\gz_{\pm}d\gz_{\pm}>=\pm 2Ddt$ with D and
$2mD=\hbar$ real one sets ${\bf (B62)}\,\,<d\gz_{\pm}d\gz_{\pm}>=\pm(D+D^*)dt$ with D
and $2mD=
\hbar=\ga+i\gb$ complex.  The complex time derivative operator becomes ${\bf (B63)}\,\,
(\gd/dt)=\pp_t+v\cdot\na-(i/2)(D+D^*)\na^2$.  Writing again ${\bf (B64)}\,\,
\psi=exp[i{\mf S}/2mD]=exp(i{\mf S}/\hbar)$ one obtains ${\bf (B65)}\,\,
v=-2iD\na log(\psi)$.  The NLSE is then obtained (via the Newton law) as ${\bf (B66)}
\,\,-\na U=m(\gd/dt)v=-2imD(\gd/dt)\na log(\psi)$.  Inserting {\bf (B63)} one gets
\bq\label{337}
\na U=2im\left[D\pp_t\na log(\psi)-2iD^2(\na log(\psi)\cdot \na)(\na log(\psi)-\frac
{i}{2}(D+D^*)D\na^2(\na log(\psi)\right]
\end{equation}
Now using the identities (i) $\na\na^2=\na^2\na$, (ii) $2(\na log(\psi)\cdot\na)
(\na log(\psi)=\na(\na log(\psi))^2$ and (iii) $\na^2log(\psi)=\na^2\psi/\psi-(\na log
(\psi))^2$ leads to a NLSE with nonlinear (kinematic pressure) potential, namely
\bq\label{338}
i\hbar\pp_t\psi=-\frac{\hbar^2}{2m}\frac{\ga}{\hbar}\na^2\psi+U\psi-i\frac{\hbar^2}{2m}
\frac{\gb}{\hbar}(\na log(\psi))^2\psi
\end{equation}
Note the crucial minus sign in front of the kinematic pressure term and also that
$\hbar=\ga+i\gb=2mD$ is complex.  When $\gb=0$ one recovers the linear SE.
The nonlinear potential is complex and one defines ${\bf (B67)}\,\,
W=-(\hbar^2/2m)(\gb/\hbar)(\na log(\psi))^2$ with U the ordinary potential; then the
NLSE is ${\bf (B68)}\,\,i\hbar\pp_t\psi=[-(\hbar^2/2m)(\ga/\hbar)
\na^2+U+iW]\psi$.  This is the fundamental result of \cite{c29}; it has the form of an
ordinary SE with complex potential $U+iW$ and complex $\hbar$.  The Hamiltonian is no
longer Hermitian and the potential itself depends on $\psi$.  Nevertheless one can
have meaningful physical solutions with real valued energies and momenta; the
homogeneity breaking hydrostatic pressure term $-b(log(\psi^*\psi)\psi$ is not present
(it would be meaningless)
and the NLSE is invariant under $\psi\to\gl \psi$.
\\[3mm]\indent
{\bf REMARK 3.5.}
One could ask why not simply propose as a valid NLSE an equation
\bq\label{339}
i\hbar\pp_t\psi=-\frac{\hbar^2}{2m}\na^2\psi+U\psi+\frac{\hbar^2}{2m}\frac{a}{m}|\na
log(\psi)|^2\psi
\end{equation}
Here one has a real Hamiltonian satisfying the homogeneity condition and the equation
admits soliton solutions of the form ${\bf (B69)}\,\,\psi=CA(x-vt)exp[i(kx-\go t)]$
where $A(x-vt)$ is to be determined by solving the NLSE.  The problem here is that 
the equation suffers from an extraneous dispersion relation.  Thus putting in the plane
wave solution $\psi\sim exp[-i(Et-px)]$ one gets an extraneous EM relation (after
setting $U=0$), namely ${\bf (B70)}\,\,E=(p^2/2m)[1+(a/m)]$ instead of the usual
$E=p^2/2m$ and hence $E_{QM}\ne E_{FT}$ where FT means field theory.$\hfill\bs$
\\[3mm]\indent
{\bf REMARK 3.6.}
It has been known since e.g. \cite{p1} that the expression for the energy functional
in nonlinear QM does not coincide with the QM energy functional, nor is it unique.
To see this write down the NLSE of \cite{b99} in the form ${\bf (B71)}\,\,
i\hbar\pp_t\psi=\pp H(\psi,\psi^*)/\pp\psi^*$ where the real Hamiltonian density is
\bq\label{340}
H(\psi,\psi^*)=-\frac{\hbar^2}{2m}\psi^*\na^2\psi+U\psi^*\psi
-b\psi^*log(\psi^*\psi)\psi+b\psi^*\psi
\end{equation}
Then using $E_{FT}=\int Hd^3r$ we see it is different from $<\hat{H}>_{QM}$ and in fact
$E_{FT}-E_{QM}=\int b\psi^*\psi d^3r=b$.  This problem does not occur in the fractal
based NLSE since it is written entirely in terms of $\psi$.$\hfill\bs$
\\[3mm]\indent
{\bf REMARK 3.7.}
In the fractal based NLSE there is no discrepancy between the QM energy functional and
the FT energy functional.  Both are given by
\bq\label{341}
N^{NLSE}_{fractal}=-\frac{\hbar^2}{2m}\frac{\ga}{\hbar}\psi^*\na^2\psi+U\psi^*\psi
-i\frac{\hbar^2}{2m}\frac{\gb}{\hbar}\psi^*(\na log(\psi)^2\psi
\end{equation}
The NLSE is unambiguously given by {\bf (B71)} and $H(\psi,\psi^*)$ is homogeneous of
degree 1 in $\gl$.  Such equations admit plane wave solutions with dispersion relation
$E=p^2/2m$; indeed, inserting the plane wave solution into the fractal based NLSE
one gets (after setting $U=0$)
\bq\label{342}
E=\frac{\hbar^2}{2m}\frac{\ga}{\hbar}\frac{p^2}{2m}+i\frac{\gb}{\hbar}\frac{p^2}{2m}
=\frac{p^2}{2m}\frac{\ga+i\gb}{\hbar}=\frac{p^2}{2m}
\end{equation}
since $\hbar=\ga+i\gb$.
The remarkable feature of the fractal approach versus all other NLSE considered sofar
is that the QM energy functional is precisely the FT one.  The complex diffusion
constant represents a truly new physical phenomenon insofar as a small imaginary
correction to the Planck constant is the hallmark of nonlinearity in QM (see
\cite{c29} for more on this).$\hfill\bs$

\section{REMARKS ON A FRACTAL SPACETIME}
\renewcommand{\theequation}{4.\arabic{equation}}
\setcounter{equation}{0}

There have been a number of articles and books involving fractal
methods in spacetime or fractal spacetime itself with impetus coming
from quantum physics and relativity.  We refer here especially to
\cite{a11,c46,c47,c40,g95,n92,n93,n94,n95,n96,n97,n98,n99,n00,
n01,n03,n04} for background to this paper.  Many related papers are
omitted here and we refer in particular to the journal Chaos,
Solitons, and Fractals CSF) for further information.
For information on fractals and stochastic processes we refer for example to
\cite{a1,b14,c33,c34,c35,f15,g16,k99,k12,l9,m11,n59,o3,p12,p16,r9,s6,t3,w2}.
We discuss here a few background ideas and constructions in order to indicate the
ingredients for El Naschie's Cantorian spacetime ${\mf E}^{\infty}$, whose exact nature
is elusive.  Suitable references are given but   
there are many more papers in the journal CSF by El Naschie
(and others)
based on these fundamental ideas and these are either important in a revolutionary sense or a
fascinating refined form of science fiction.  In what appears at times to be pure numerology
one manages to (rather hastily) produce amazingly close numerical approximations to virtually
all the fundamental constants of physics (including string theory).  The key concepts
revolve around the famous golden ratio
$(\sqrt{5}-1)/2$ and a strange Cantorian space ${\mf E}^{\infty}$ which we try to describe
below.  It is very tempting to want all of these (heuristic) results to be true and the
approach seems close enough and universal enough to compel one to think something very
important must be involved.  Moreover such scope and accuracy cannot be ignored so we try to 
examine some of the constructions in a didactic manner in order to possibly generate 
some understanding.

\subsection{COMMENTS ON CANTOR SETS}

\begin{example}
In the paper \cite{m11} one discusses random recursive constructions leading to Cantor sets,
etc.  Associated with each such construction is a universal number $\ga$ such that almost
surely the random object has Hausdorff dimension $\ga$ (we assume that ideas of Hausdorff and
Minkowski-Bouligand (MB) or upper box dimension are known - cf. \cite{b14,c46,f15,l9}).
One construction of a Cantor set goes as follows.  Choose $x$ from $[0,1]$ according to the
uniform distribution and then choose $y$ from $[x,1]$ according to the uniform distribution
on $[x,1]$.  Set $J_0=[0,x]$ and $J_1=[y,1]$ and recall the standard $1/3$ construction for
Cantor sets.  Continue this procedure by rescaling to each of the intervals already obtained.
With probability one one then obtains a Cantor set $S_c^0$ with Hausdorff dimension
${\bf (C1)}\,\,\ga=\phi= (\sqrt{5}-1)/2\sim .618$.  Note that this is just a particular
random Cantor set; there are others with different Hausdorff dimensions (there seems to be
some - possibly harmless - confusion on this point in the El Naschie papers).  However the
golden ratio $\phi$ is a very interesting number whose importance rivals that of $\pi$ or
$e$.  In particular (cf.
\cite{a1}) $\phi$ is the hardest number to approximate by rational numbers and could be
called the most irrational number.  This is because its continued fraction represention
involves all $1's$.
$\hfill\bs$
\end{example}
\begin{example}
From \cite{n92} the Hausdorff (H) dimension of a traditional triadic Cantor set is $d_c^{(0)}
=log(2)/log(3)$.  To determine the equivalent to a triadic Cantor set in 2 dimensions one
looks for a set which is triadic Cantorian in all directions.  The analogue of an area
$A=1\times 1$ is a quasi-area $A_c=d_c^{(0)}\times d_c^{(0)}$ and to normalize $A_c$ one uses
$\rho_2=(A/A_c)_2=1/(d_c^{(0)})^2$ (for n-dimensions ${\bf
(C2)}\,\,\rho_n=1/(d_c^{(0)})^{n-1}$).  Then the $n^{th}$ Cantor like H dimension $d_c^{(n)}$
will have the form ${\bf (C3)}\,\,d_c^{(n)}=\rho_nd_c^{(0)}=1/(d_c^{(0)})^{n-1}$.  Note also
that the H dimension of a Sierpinski gasket is ${\bf (C4)}\,\,d_c^{(n+1)}/d_c^{(n)}=
1/d_c^{(0)}=log(3)/log(2)$ and in any event the straight-forward interpretation of
$d_c^{(2)}=log(3)/log(2)$ is a scaling of $d_c^{(0)}=log(2)/log(3)$ proportional to the ratio
of areas $(A/A_c)_2$.  One notes that ${\bf
(C5)}\,\,d_c^{(4)}=1/(d_c^{(0)})^3=(log(3)/log(2))^3
\simeq 3.997\sim 4$ so the 4-dimensional Cantor set is essentially ``space filling".
\\[3mm]\indent
Another derivation goes as follows.  Define probability quotients
$\gO=dim(subset)/dim(set)$.  For a triadic Cantor set in 1-D ${\bf (C6)}\,\,\gO^{(1)}=
d_c^{(0)}/d_c^{(1)}=d_c^{(0)}\,\,(d_c^{(1)}=1)$.  To lift the Cantor set to n-dimensions look
at the multiplicative probability law ${\bf (C7)}\,\,\gO^{(n)}=(\gO^{(1)})^n=(d_c^{(0)})^n$.
However since $\gO^{(1)}=d_c^{(0)}/d_c^{(n)}$ we get ${\bf (C8)}\,\,d_c^{(0)}/d_c^{(n)}=
(d_c^{(0)})^n\Rightarrow d_c^{(n)}=1/(d_c^{(0)})^{n-1}$.  Since $\gO^{(n-1)}$ is the
probability of finding a Cantor point (Cantorian) one can think of the H dimension
$d_c^{(n)}=1/\gO^{(n-1)}$ as a measure of ignorance.  One notes here also that for $d_c^{(0)}
=\phi$ (the Cantor set $S_c^{(0)}$ of Example 2.1) one has $d_c^{(4)}=1/\phi^3=4+\phi^3
\simeq 4.236$ which is surely space filling.$\hfill\bs$ 
\end{example}
\indent
Based on these ideas one proves in \cite{n93,n94,n96} a number of theorems and we sketch some
of this here.  One picks a ``backbone" Cantor set with H dimension $d_c^{(0)}$ (the choice
of $\phi=d_c^{(0)}$ will turn out to be optimal for many arguments).  Then one imagines a
Cantorian spacetime ${\mf E}^{\infty}$ built up of an infinite number of spaces of dimension
$d_c^{(n)}\,\,(-\infty\leq n <\infty)$.  The exact form of embedding etc. here is not
specified so one imagines e.g. ${\mf E}^{\infty}=\cup {\mf E}^{(n)}$ 
(with unions and intersections) in some amorphous sense.
There are some connections of this to vonNeumann's continuous geometries indicated in
\cite{n98}.  In this connection we remark that only ${\mf E}^{(-\infty)}$ is the completely
empty set (${\mf E}^{(-1)}$ is not empty).  First we note that $\phi^2+\phi-1=0$ leading to
${\bf (C9)}\,\,1+\phi=1/\phi,\,\,\phi^3=(2+\phi)/\phi,\,\,(1+\phi)/(1-\phi)=1/\phi(1-\phi)=
4+\phi^3=1/\phi^3$ (a very interesting number indeed).  Then one asserts that
\begin{theorem}
Let $(\gO^{(1)})^n$ be a geometrical measure in n-dimensional space of a multiplicative point
set process and $\gO^{(1)}$ be the Hausdorff dimension of the backbone (generating) set
$d_c^{(0)}$.  Then $<d>=1/d_c^{(0)}(1-d_c^{(0)})$ (called curiously an average Hausdorff
dimension) will be exactly equal to the average space dimension
${}_{\widetilde{}}<n>=(1+d_c^{(0)})(1-d_c^{(0)})$ and equivalent to a 4-dimensional Cantor
set with H-dimension $d_c^{(4)}=1/(d_c^{(0)})^3$ if and only if 
$d_c^{(0)}=\phi$.
\end{theorem}
\indent
To see this take $\gO^{(n)}=(\gO^{(1)})^n$ again and consider the total probability of the
additive set described by the $\gO^{(n)}$, namely
${\bf (C10)}\,\,Z_0=\sum_0^{\infty}(\gO^{(1)})^n=1/(1-\gO^{(1)})$.  It is conceptually easier
here to regard this as a sum of weighted dimensions (since $d_c^{(n)}=1/(d_c^{(0)})^{n-1}$)
and consider $w_n=n(d_c^{(0)})^n$.  Then the expectation of $n$ becomes (note $d_c^{(n)}\sim
1/(d_c^{(0)})^{n-1}\sim 1/\gO^{(n-1)}$ so $n(d_c^{(0)})^{n-1}\sim n/d_c^{(n)}$)
\bq\label{41}
E(n)=\frac{\sum_1^{\infty}n^2(d_c^{(0)})^{n-1}}{\sum_1^{\infty}n(d_c^{(0)})^{n-1}}=
{}_{\widetilde{}}<n>=\frac{1+d_c^{(0)}}{1-d_c^{(0)}}
\end{equation}
Another average here is defined via (blackbody gamma distribution)
\bq\label{42}
<n>=\frac{\int_0^{\infty}n^2(\gO^{(1)})^ndn}{\int_0^{\infty}n(\gO^{(1)})^ndn}=\frac{-2}
{log(\gO^{(1)})}
\end{equation}
which corresponds to ${}_{\widetilde{}}<n>$ after expanding the logarithm and omitting higher
order terms.  However ${}_{\widetilde{}}<n>$ seems to be the more valid calculation
here.  Similarly one defines (somewhat ambiguously) an expected value for
$d_c^{(n)}$ via
\bq\label{43}
<d>=\frac{\sum_1^{\infty}n(d_c^{(0)})^{n-1}}{\sum_1^{\infty}(d_c^{(0)})^n}=
\frac{1}{d_c^{(0)}(1-d_c^{(0)})}
\end{equation}
This is contrived of course (and cannot represent $E(d_c^{(n)})$ since one is computing
reciprocals $\sum (n/d_c^{(n)})$ but 
we could think of computing an expected ignorance and identifying this with the
reciprocal of dimension.  Thus
the label $<d>$ does not seem to represent an expected dimension but if we accept it as a
symbol then for
$d_c^{(0)}=\phi$ one has from {\bf (C9)}
\bq\label{44}
{}_{\widetilde{}}<n>=\frac{1+\phi}{1-\phi}=<d>=
\frac{1}{\phi(1-\phi)}=d_c^{(4)}=4+\phi^3=\frac{1}{\phi^3}\sim 4.236
\end{equation}
\indent
{\bf REMARK 4.1.}
We note that the normalized probability ${\bf
(C11)}\,\,N=\gO^{(1)}/Z_0=\gO^{(1)}(1-\gO^{(1)})= 1/<d>$ for any $d_c^{(0)}$.  Further if
$<d>=4=1/d_c^{(0)}(1-d_c^{(0)})$ one has $d_c^{(0)}=1/2$ while
${}_{\widetilde{}}<n>=3<4=<d>$.  One sees also that $d_c^{(0)}=1/2$ is the minimum (where
$d<d>/d(d_c^{(0)})=0$).$\hfill\bs$
\\[3mm]\indent
{\bf REMARK 4.2.}
The results of Theorem 4.1 should really be phrased in terms of ${\mf E}^{\infty}$ (cf.
\cite{n99}). thus ($H\sim$ Hausdorff dimension and $T\sim$ topological dimension)
\bq\label{45}
dim_H{\mf E}^{(n)}=d_c^{(n)}=\frac{1}{(d_c^{(0)})^{n-1}};
\end{equation}
$$<d>=\frac{1}{d_c^{(0)}(1-d_c^{(0)})};\,\,{}_{\widetilde{}}<dim_T{\mf
E}^{\infty}>=\frac{1+d_c^{(0)}}{1-d_c^{(0)}}={}_{\widetilde{}}<n>$$
In any event ${\mf E}^{\infty}$ is formally infinite dimensional but effectively it is $4\,\pm$
dimensional with an infinite number of internal dimensions.  We emphasize that ${\mf E}^{\infty}$
appears to be constructed  
from a fixed backbone Cantor set with H dimension
$1/2\leq d_c^{(0)}< 1$; thus each such $d_c^{(0)}$ generates an ${\mf E}^{\infty}$ space.
Note that in \cite{n99} ${\mf E}^{\infty}$ is looked upon as a transfinite discretum (?)
underpinning the continuum.  
$\hfill\bs$
\\[3mm]\indent
{\bf REMARK 4.3.}
An interesting argument from \cite{n98} goes as follows.  Thinking of $d_c^{(0)}$ as a geometrical
probability one could say that the spatial (3-dimensional) probability of finding a
Cantorian ``point" in ${\mf E}^{\infty}$ must be given by the intersection probability
${\bf (C12)}\,\, P=(d_c^{(0)})^3$ where $3\sim$ 3 topological spatial dimension.  P could
then be regarded as a Hurst exponent (cf. \cite{a11,n5,w2}) and the Hausdorff dimension
of the fractal path of a Cantorian would be ${\bf
(C13)}\,\,d_{path}=1/H=1/P=1/(d_c^{(0)})^3$.  Given $d_c^{(0)}=\phi$ this means 
$d_{path}=4+\phi^3\sim 4^{+}$ so a Cantorian in 3-D would sweep out a 4-D world sheet; i.e. the
time dimension is created by the Cantorian space ${\mf E}^{\infty}$ (! - ?).  Conjecturing
further (wildly) one could say that perhaps space (and gravity) is created by the fractality
of time. This is a typical form of conjecture to be found in the El Naschie papers - extremely
thought provoking but ultimately heuristic.  Regarding the Hurst exponent one recalls that for
Feynmann trajectories in $1+1$ dimensions ${\bf
(C14)}\,\,d_{path}=1/H=1/d_c^{(0)}=d_c^{(2)}$.  Thus we are concerned with relating {\bf
(C13)} and {\bf (C14)} (among other matters).  Note that path dimension is often thought of
as a fractal dimension (M-B or box dimension), which is not necessarily the same as the
Hausdorff dimension.  However in \cite{a11} one shows that quantum mechanical free motion
produces fractal paths of Hausdorff dimension 2 (cf. also \cite{k14}).
$\hfill\bs$
\\[3mm]\indent
{\bf REMARK 4.4.}
Following \cite{c41} let $S_c^{(0)}$ correspond to the set with dimension $d_c^{(0)}=\phi$. 
Then the complementary dimension is $\tl{d}_c^{(0)}=1-\phi=\phi^2$.  The path dimension
is gien as in {\bf (C14)} by ${\bf (C15)}\,\,d_{path}=d_c^{(2)}=1/\phi=1+\phi$ and 
$\tl{d}_{path}=\tl{d}_c^{(2)}=1/(1-\phi)=1/\phi^2=(1+\phi)^2$.  Following El Naschie for an
equivalence between unions and intersections in a given space one requires (in the present
situation) that
\bq\label{47}
d_{crit}=d_c^{(2)}+\tl{d}_c^{(2)}=\frac{1}{\phi}+\frac{1}{\phi^2}=\frac{\phi(1+\phi)}{\phi^3}
=\frac{1}{\phi^3}=\frac{1}{\phi}\cdot\frac{1}{\phi^2}=d_c^{(2)}\cdot\tl{d}_c^{(2)}=4+\phi^3
\end{equation}
where ${\bf (C16)}\,\,d_{crit}=4+\phi^3=d_c^{(4)}\sim 4.236$.  Thus the critical dimension
coincides with the Hausdorff dimension of $S_c^{(4)}$ which is embedded densely into a smooth
space of topological dimension 4.  On the other hand the backbone set of dimension
$d_c^{(0)}=\phi$ is embedded densely into a set of topological dimension zero (a point). 
Thus one thinks in general of $d_c^{(n)}$ as the H dimension of a Cantor set of dimension
$\phi$ embedded into a smooth space of integer topological dimension n.$\hfill\bs$
\\[3mm]\indent
{\bf REMARK 4.5.}
In \cite{c41} it is also shown that realization of the spaces ${\mf E}^{(n)}$ comprising 
${\mf E}^{\infty}$ can be expressed via the fractal sprays of Lapidus-van Frankenhuysen
(cf. \cite{l9}).  Thus we refer to \cite{l9} for graphics and details and simply sketch some
ideas here (with apologies to M. Lapidus).  A fractal string is a bounded open subset of {\bf
R} which is a disjoint union of an infinite number of open intervals ${\mf
L}=\ell_1,\ell_2,\cdots$.  The geometric zeta function of ${\mf L}$ is ${\bf (C17)}\,\,
\gz_{{\mf L}}(s)=\sum_1^{\infty}\ell_j^{-s}$.  One assumes a suitable meromorphic extension
of $\gz_{{\mf L}}$ and the complex dimensions of ${\mf L}$ are defined as the poles of this
meromorphic extension.  The spectrum of ${\mf L}$ is the sequence of frequencies $f=k\cdot
\ell_j^{-1}\,\,(k=1,2,\cdots)$ and the spectral zeta function of ${\mf L}$ is defined as 
${\bf (C18)}\,\,\gz_{\nu}(s)=\sum_ff^{-s}$ where in fact $\gz_{\nu}(s)=\gz_{{\mf L}}(s)
\gz(s)$ (with $\gz(s)$ the classical Riemann zeta function).  Fractal sprays are higher
dimensional generalizations of fractal strings.  As an example consider the spray $\gO$
obtained by scaling an open square B of size 1 by the lengths of the standard triadic Cantor
string CS.  Thus $\gO$ consists of one open square of size 1/3, 2 open squares of size 1/9,
4 open squares of size 1/27, etc. (see \cite{l9} for pictures and explanations).  Then the
spectral zeta function for the Dirichlet Laplacian on the square is ${\bf
(C19)}\,\,\gz_B(s)=\sum_{n_1,n_2=1}^{\infty}(n_1^2+n_2^2)^{s/2}$ and the spectral zeta
function of the spray is ${\bf (C20)}\,\,\gz_{\nu}(s)=\gz_{CS}(s)\cdot \gz_B(s)$.  Now 
${\mf E}^{\infty}$ is composed of an infinite hierarchy of sets ${\mf E}^{(j)}$ with dimension
$(1+\phi)^{j-1}=1/\phi^{j-1}\,\,(j=0,\pm 1,\pm 2,\cdots)$ and these sets correspond to a
special case of boundaries $\pp\gO$ for fractal sprays $\gO$ 
whose scaling ratios are suitable
binary powers of $2^{-\phi^{j-1}}$. Indeed for $n=2$ the spectral zeta function of the fractal
golden spray indicated above is ${\bf (C21)}\,\,\gz_{\nu}(s)=(1/(1-2\cdot
2^{s\phi})\gz_B(s)$.  The poles of
$\gz_B(s)$ do not coincide with the zeros of the denominator $1-2\cdot 2^{-s\phi}$ so the
(complex) dimensions of the spray correspond to those of the boundary $\pp\gO$ of $\gO$.  One
finds that the real part $\Re s$ of the complex dimensions coincides with $dim\,{\mf
E}^{(2)}=1+\phi= 1/\phi^2$ and one identifies then $\pp\gO$ with ${\mf E}^{(2)}$.  The
procedure generalizes to higher dimensions (with some stipulations) and for dimension n there
results $\Re s= 1/\phi^{n-1}=dim\,{\mf E}^{(n)}$.  This produces a physical model of the
Cantorian fractal space from the boundaries of fractal sprays (see \cite{c41} for further
details and \cite{l9} for precision).  Other (putative) geometric realizations of ${\mf
E}^{\infty}$ are indicated in \cite{n02} in terms of wild topologies, etc.
$\hfill\bs$

\section{HYDRODYNAMICS AND THE FRACTAL SCHR\"ODINGER EQUATION}
\renewcommand{\theequation}{5.\arabic{equation}}
\setcounter{equation}{0}

We sketch first some material from \cite{a16} 
(see also \cite{c46,n5,n13,n30} and Sections 2-4 for background).  Thus let $\psi$ be the
wave function of a test particle of mass
$m_0$ in a force field $U(r,t)$ determined via ${\bf
(D1)}\,\,i\hbar\pp_t\psi=U\psi-(\hbar^2/2m)\na^2\psi$ where $\na^2=\gD$.  One writes
${\bf (D2)}\,\,\psi(r,t)=R(r,t)exp(iS(r,t))$ with ${\bf (D3)}\,\,v=(\hbar/2m)\na S$ and
$\rho=R\cdot R$ (one assumes $\rho\ne 0$ for physical meaning).  Thus the field
equations of QM in the hydrodynamic picture are 
\bq\label{51}
d_t(m_0\rho v)=\pp_t(m_0\rho v)+\na (m_0\rho
v)=-\rho\na(U+Q);\,\,\pp_t\rho+\na\cdot(\rho v)=0
\end{equation}
where ${\bf (D4)}\,\,Q=-(\hbar^2/2m_0)(\gD\sqrt{\rho}/\sqrt{\rho})$ is the quantum
potential (or interior potential).  Now because of the nondifferentiability of spacetime
an infinity of geodesics will exist between any couple of points A and B.  The ensemble
will define the probability amplitude (this is a nice assumption but what is a geodesic here). 
At each intermediate point C one can consider the family of incoming (backward) and outgoing
(forward) geodesics and define average velocities $b_{+}(C)$ and $b_{-}(C)$ on these
families.  These will be different in general and following Nottale
this doubling of the velocity vector is at the origin of the complex nature of QM.
Even though Nottale reformulates Nelson's stochastic QM the former's interpretation is
profoundly different.  While Nelson (cf. \cite{n6}) assumes an underlying Brownian
motion of unknown origin which acts on particles in a still Minkowskian spacetime,
and then introduces nondifferentiability as a byproduct of this hypothesis, Nottale
assumes as a fundamental and universal principle that spacetime itself is no longer
Minkowskian nor differentiable.  While with Nelson's Browian motion hypothesis,
nondifferentiability is but an approximation which expected to break down at the scale
of the underlying collisions (?), where a new physics should be introduced, Nottale's
hypothesis of nondifferentiability is essential and should hold down to the smallest
possible length scales.  (This sentence is interesting but needs elaboration).
Following Nelson one defines now the mean forward and backward derivatives
\bq\label{52}
\frac{d_{\pm}}{dt}y(t)=lim_{\gD t\to 0_{\pm}}\left<\frac{y(t+\gD t)-y(t)}{\gD t}\right>
\end{equation}
This gives forward and backward mean velocities
${\bf (D5)}\,\,(d_{+}/dt)x(t)=b_{+}$ and $(d_{-}/dt)x(t)=b_{-}$ for a position vector x. 
Now in Nelson's stochastic mechanics one writes two systems of equations for the forward
and backward processes and combines them in the end in a complex equation, Nottale works
from the beginning with a complex derivative operator 
\bq\label{53}
\frac{\gd}{dt}=\frac{(d_{+}+d_{-})-i(d_{+}-d_{-})}{2dt}
\end{equation}
leading to ${\bf (D6)}\,\,V=(\gd/dt)x(t)=v-iu=(1/2)(b_{+}+b_{-})-(i/2)(b_{+}-b_{-})$.
One defines also ${\bf (D7)}\,\,(d_v/dt)=(1/2)(d_{+}+d_{-})/dt$ and $(d_u/dt)=
(1/2)(d_{+}-d_{-})/dt$ so that $d_vx/dt=v$ and $d_ux/dt=u$.  Here $v$ generalizes the
classical velocity while $u$ is a new quantity arising from nondifferentiability.  This
leads to a stochastic process satisfying (respectively for the forward $(dt>0)$ and
backward $(dt<0)$ processes) ${\bf (D8)}\,\,dx(t)=b_{+}[x(t)]+d\xi_{+}(t)=
b_{-}[x(t)]+d\xi_{-}(t)$.  The $d\xi(t)$ terms can be seen as fractal functions and they
amount to a Wiener process when ${\mc D}=2$ (presumably the fractal dimension).  Then
the $d\xi(t)$ are Gaussian with mean zero, mutually independent, and satisfy
${\bf (D9)}\,\,<d\xi_{\pm i}d\xi_{\pm j}>=\pm2D\gd_{ij}dt$ where D is a diffusion
coefficient.  D can be found via $D=\hbar/2m_0$ given $\tau_0=\hbar/(m_0c^2)$ (deBroglie
time scale in the rest frame - cf \cite{c46} for more on this).  Now {\bf (D9)} allows
one to give a general expression for the complex time derivative, namely 
\bq\label{54}
df=\frac{\pp f}{\pp t}+\na f\cdot dx+\frac{1}{2}\frac{\pp^2f}{\pp x_i\pp x_j}dx_idx_j
\end{equation}
Next compute the forward and backward derivatives of $f$.  Then $<dx_idx_j>\to
<d\xi_{\pm i}d\xi_{\pm j}>$ so the last term in \eqref{54} amounts to a Laplacian 
via {\bf (D9)} and one obtains ${\bf
(D10)}\,\,(d_{\pm}f/dt)=[\pp_t+b_{\pm}\cdot\na\pm D\gD]f$.  This is an important
result. Thus assume the fractal dimension is not 2 in which case there is no longer a
cancellation of the scale dependent terms in \eqref{54} and instead of $D\gD f$  one
would obtain an explicitly scale dependent behavior $D\gd t^{(2/D)-1}\gD f$.  In other
words the value $D=2$ implies that the scale symmetry becomes hidden in the operator
formalism.  Using {\bf (D10)} one obtains the complex time derivative operator in the
form ${\bf (D11)}\,\,(\gd/dt)=\pp_t+V\cdot\na-iD\gD$ (cf. {\bf (D6)} for V).
Nottale's prescription is then to replace $d/dt$ by $\gd/dt$.  In this spirit one can
write now ${\bf (D12)}\,\,\psi=exp(i(\mf S/2m_0D))$ so that ${\bf (D13)}\,\,
V=-2iD\na (log(\psi))$ and then the generalized Newton equation ${\bf (D14)}\,\,-\na
U=m_0(\gd/dt)V$ reduces to the SE.
\\[3mm]\indent
Now assume the velocity field from the hydrodynamic model agrees with the real part $v$ of
the complex velocity V and equate the wave functions from the two models {\bf (D12)} and 
{\bf (D2)}; one obtains for ${\mf S}=s+i\gs\,\,\,{\bf
(D15)}\,\,s=2m_0DS,\,\,D=(\hbar/2m_0),$ and $\gs=-m_0Dlog(\rho)$.  Using the definition
${\bf (D16)}\,\,V=(1/m_0)\na{\mf S}=(1/m_0)\na s+(i/m_0)\na\gs=v-iu$ (which results via
{\bf (D6)} by putting {\bf (D12)} into {\bf (D13)}) we get
${\bf (D17)}\,\,v=(1/m_0)\na s=2D\na S$ and $u=-(1/m_0)\na\gs=D\na log(\rho)$.  Note that
the imaginary part of the complex velocity given in {\bf (D17)} coincides with Nottale.
Dividing the time dependent SE {\bf (D1)} by $2m_0$ and taking the gradient gives
${\bf (D18)}\,\,\na U/m_0=2D\na[i\pp_tlog(\psi)+D(\gD \psi/\psi)]$ where $\hbar/2m_0$ has
been replaced by D.  Then consider the identities
\bq\label{55}
\gD \na=\na \gD;\,\,(\na f\cdot\na)(\na f)=(1/2)\na(\na f)^2;\,\,\frac{\gD f}{f}=\gD
log(f)+(\na log(f))^2
\end{equation}
Now the second term in the right of {\bf (D18)} becomes ${\bf
(D19)}\,\,\na(\gD\psi/\psi)=\gD(\na log(\psi))+2(\na log(\psi)\cdot\na)(\na log(\psi))$
so {\bf (D18)} can be written as ${\bf (D20)}\,\,\na U=2iDm_0[\pp_t\na
log(\psi)-iD\gD(\na log(\psi)-2iD(\na log(\psi)\cdot\na)(\na log(\psi))]$.
One can show that {\bf (D20)} is nothing but the generalized Newton equation {\bf
(D14)}.  Now if we replace the complex velocity {\bf (D13)}, taking into account
{\bf (D6)} and {\bf (D17)} we get
\bq\label{56}
-\na U=m_0\{\pp_t(v-iD\na log(\rho)+[i(v-iD\na log(\rho)\cdot\na](v-iD\na
log(\rho))-
\end{equation}
$$-iD\gD(v-iD\na log(\rho))\}$$
Equation \eqref{56} is a complex differential equation and reduces to (using
\eqref{55})
\bq\label{57}
m_0[\pp_tv+(v\cdot\na)v]=-\na\left(U-2m_0D^2\frac{\gD\sqrt{\rho}}{\sqrt{\rho}}\right);\,\,
\na\left\{\frac{1}{\rho}\left[\pp_t\rho+\na\cdot(\rho v)\right]\right\}
\end{equation}
The last equation in \eqref{57} reduces to the continuity equation up to a phase factor
$\ga(t)$ which can be set equal to zero (note again that $\rho\ne 0$ is posited).  Thus
\eqref{57} is nothing but the fundamental equations \eqref{51} of the hydrodynamic
model.  Further combining the imaginary part of the complex velocity in {\bf (D17)}
with the quantum potential {\bf (D4)} and using \eqref{55} one gets ${\bf
(D21)}\,\,Q=-m_0D\na\cdot u-(1/2)m_0u^2$.  Since u arises from nondifferentiability
according to our nondifferentiable space model of QM it follows that the quantum
potential comes from the nondifferentiability of the quantum spacetime (very nice but 
where is ${\mf E}^{\infty}$ from the title of \cite{a16} - also the $x$ derivatives should
be clarified).
\\[3mm]\indent
Putting $U=0$ in the first equation of \eqref{57}, multiplying by $\rho$, and taking
the second equation into account yields
\bq\label{58}
\pp_t(m_0\rho \nu_k)+\frac{\pp}{\pp x_i}(m_0\rho\nu_i\nu_k)=-\rho\frac{\pp}{x_k}\left[
2m_0D^2\frac{1}{\sqrt{\rho}}\frac{\pp}{\pp x_i}\frac{\pp}{\pp x_i}(\sqrt{\rho})\right]
\end{equation}
(here $\nu_k\sim v_k$ seems indicated).  Now set ${\bf
(D22)}\,\,\Pi_{ik}=m_0\rho\nu_i\nu_k-\gs_{ik}$ along with $\gs_{ik}=m_0\rho D^2(\pp/\pp
x_i)(\pp/\pp x_k)(log(\rho))$.  Then \eqref{58} takes the simple form ${\bf (D23)}\,\,
\pp_t(m_0\rho\nu_k)=-\pp\Pi_{ik}/\pp x_i$.  The analogy with classical fluid
mechanics works well if one introduces the kinematic ${\bf (D24)}\,\,\mu=D/2$ and
dynamic
$\eta=(1/2)m_0D\rho$ viscosities.  Then $\Pi_{ik}$ defines the momentum flux density
tensor and $\gs_{ik}$ the internal stress tensor ${\bf (D25)}\,\,\gs_{ik}=\eta[(\pp
u_i/\pp x_k)+(\pp u_k/\pp x_i)]$.  From {\bf (D22)} one can see that the internal stress
tensor is build up using the quantum potential while the equations \eqref{51} or
\eqref{57} are nothing but systems of Navier-Stokes type for the motion where the
quantum potential plays the role of an internal stress tensor.  In other words the
nondifferentiability of the quantum spacetime manifests itself like an internal stress
tensor.  For clarity in understanding {\bf (D23)} we put this in one dimensional form so
\eqref{58} becomes
\bq\label{59}
\pp_t(m_0\rho v)+\pp_x(m_0\rho v^2)=-\rho\pp\left(2m_0D^2\frac{1}{\sqrt{\rho}}
\pp^2\sqrt{\rho}\right)=\rho\pp Q
\end{equation}
and $\Pi=m_0\rho v^2-\gs$ with $\gs=m_0\rho D^2\pp^2log(\rho)$.  This agrees in the
standard formulas (cf. \cite{c46}).  Now note
$\pp\sqrt{\rho}=(1/2)\rho^{-1/2}\rho'$
and $\pp^2\sqrt{\rho}=(1/2)[-(1/2)\rho^{-3/2}(\rho')^2+
\rho^{-1/2}\rho'']$ with $\pp^2log(\rho)=\pp(\rho'/\rho)=(\rho''/\rho)-(\rho'/\rho)^2$
while
\bq\label{510}
-\rho\pp\left[2m_0D^2\frac{1}{\sqrt{\rho}}\left(\pp^2\sqrt{\rho}\right)\right]=
-2m_0D^2\rho\pp\left[\frac{1}{2\sqrt{\rho}}\left(-\frac{1}{2}\rho^{-3/2}(\rho')^2
+\rho^{-1/2}\rho''\right)\right]=
\end{equation}
$$=-2m_0D^2\rho\pp\left[\frac{\rho''}{2\rho}-\frac{1}{4}\left(\frac{(\rho'}{\rho}\right)^2\right]
=-m_0D^2\rho\pp\left[\frac{\rho''}{\rho}
-\frac{1}{2}\left(\frac{\rho'}{\rho}\right)^2\right]$$
One wants to show then that {\bf (D23)} holds or equivalently $-\pp\gs=\eqref{510}$. 
Here
\bq\label{511}
-\pp\gs=-\pp[m_0\rho
D^2\pp^2log(\rho)]=-m_0D^2\left[\rho'\left(\frac{\rho''}{\rho}-\left(\frac{\rho'}{\rho}
\right)^2\right)+\rho\pp\left(\frac{\rho''}{\rho}-\frac{(\rho')^2}{\rho}\right)\right]
\end{equation}
so we want $\eqref{511}=\eqref{510}$ and this is easily verified.

\section{RECAPITULATION}
\renewcommand{\theequation}{6.\arabic{equation}}
\setcounter{equation}{0}

We write down now some of the main formulas here (with some unification of notation) in
order to help provide perspective.  The goal is not entirely clear but many questions will
arise as we go along and at the end.  Hopefully we will be able to answer some of the
questions.
\begin{enumerate}
\item
We write from Section 2 ${\bf (E1)}\,\,\psi=Rexp(iS/\hbar)$ with
\bq\label{61}
S_t+\frac{(S')^2}{2m}+V-\frac{\hbar^2}{2m}\frac{R''}{R}=0;\,\,\pp_t(R^2)+\frac{1}{m}
(R^2S')'=0
\end{equation}
For $P=R^2$ and $Q=-(\hbar^2/2m)(R''/R)$ this yields
\bq\label{62}
S_t+\frac{(S')^2}{2m}+Q+V=0;\,\,P_t+\frac{1}{m}(PS')'=0
\end{equation}
Writing $\rho=mP$ and $p=m\dot{x}$ leads to
\bq\label{63}
\pp_t(\rho v)+\pp(\rho v^2)+\frac{\rho}{m}\pp V-\frac{\hbar^2}{2m^2}\rho\pp\left(
\frac{\pp^2\sqrt{\rho}}{\sqrt{\rho}}\right)=0
\end{equation}
Along the way one arrived at \eqref{2.8} and ``completed" this with a pressure term
$\na F=\rho^{-1}\na{\mf p}$ or $F'=(1/R^2){\mf p}'$ to arrive at ${\bf (E2)}\,\,
mv_t+mvv'=-\pp(V+Q)-F'$ corresponding to a SE ${\bf (E3)}\,\,i\hbar\psi_t=-(\hbar^2/2m)
\psi''+V\psi+F\psi$.  One wants then $F=F(\psi)$.
\item
Consider a quantum state corresponding to a ``subquantum" statistical ensemble governed by
classical kinetics in a phase space.  One arrives at $\psi=\rho^{1/2}exp(i{\mf S}/\hbar)$
with ${\bf (E4)}\,\,i\hbar\psi_t=-(\hbar^2/2m)\psi_{xx}+{\mc V}\psi$ where ${\mf S}=NS,\,\,N=
\int|\psi|^2d^nx,\,\,\hbar=N\eta,\,\,m=N\mu,\,\,{\mc
V}=NV,$ and $log(\psi)=(1/2)log(\rho)+(i/\eta)S$.  The fields $\rho,\,S$ or $\xi,\,S$
determine a quantum fluid with (cf. \eqref{2.33})
\bq\label{64}
\frac{\pp\xi}{\pp t}+\frac{1}{\mu}\frac{\pp^2S}{\pp x^2}+\frac{1}{\mu}\frac{\pp\xi}
{\pp x}\frac{\pp S}{\pp x}=0;
\end{equation}
$$\frac{\pp S}{\pp t}-\frac{\eta^2}{4\mu}\frac{\pp^2\xi}{\pp
x^2}-\frac{\eta^2}{8\mu}\left(\frac{\pp \xi}{\pp
x}\right)^2+\frac{1}{2\mu}\left(\frac{\pp S}{\pp x}\right)^2+V=0$$
which for $\psi=\rho^{1/2}exp(i{\mf S}/\hbar)$ leads to
\bq\label{644}
i\hbar\frac{\pp\Psi}{\pp t}=-\frac{\hbar^2}{2m}\frac{\pp^2 \Psi}{\pp x^2}+{\mc
V}\Psi
\end{equation}
\item
The Fisher information connection \a la Remarks 2.4-2.5 involves a classical ensemble with
particle mass m moving under a potential V
\bq\label{65}
S_t+\frac{1}{2m}(S')^2+V=0;\,\,P_t+\frac{1}{m}\pp(PS')'=0
\end{equation}
where S is a momentum potential; note that no quantum potential is present but this will
be added on in the form of a term $(1/2m)\int dt(\gD N)^2$ in the Lagrangian which measures
the strength of fluctuations.  This can then be specified in terms of the probability
density P as indicated in Remark 2.4 leading to a SE \eqref{5.21}.  A ``neater" approach
is given in Remark 2.5 leading in 1-D to
\bq\label{65}
S_t+\frac{1}{2m}(S')^2+V+\frac{\gl}{m}\left(\frac{(P')^2}{P^2}-\frac{2P''}{P}\right)=0
\end{equation}
Note that $Q=-(\hbar^2/2m)(R''/R)$ becomes for $R=P^{1/2}$ ${\bf (E5)}\,\,
Q=-(2\hbar^2/2m)[(2P''/P)-(P'/P)^2]$ (cf. \eqref{2.4}).  Thus the addition of the Fisher
information serves to quantize the classical system.
\item
One defines an information entropy (IE) in Remark 2.6 via ${\bf (E6)}\,\,{\mf S}=-\int \rho
log(\rho)d^3x\,\,(\rho=|\psi|^2)$ leading to 
\bq\label{67}
\frac{\pp{\mf S}}{\pp t}=\int (1+log(\rho))\pp(v\rho)\sim \int \frac{(\rho')^2}{\rho}
\end{equation}
modulo constants involving $D\sim \hbar/2m$.  ${\mf S}$ is typically not conserved and 
$\pp_t\rho=-\na\cdot(v\rho)\,\,(u=D\na log(\rho)$ with $v=-u$ corresponds to standard
Brownian motion with
$d{\mf S}/dt\geq 0$.  Then high IE production corresponds to rapid flattening of the
probability density.  Note here also that ${\mf F}\sim -(2/D^2)\int \rho Qdx=\int
dx[(\rho')^2/\rho]$ is a functional form of Fisher information.  Entropy balance is
discussed in Remark 2.8 and the manner in which Q appears in the hydrodynamical formalism
is exhibited in \eqref{5.59}-\eqref{5.60}.
\item
The Nagasawa theory (based in part on Nelson's work) is very revealing and fascinating
(see \cite{n9,n10}).  The essense of Theorem 3.1 is that $\psi=exp(R+iS)$ satisfies the SE
${\bf (E7)}\,\,i\psi_t+(1/2)\psi''+ia\psi'-V\psi=0$ if and only if
\bq\label{68}
V=-S_t+\frac{1}{2}R''+\frac{1}{2}(R')^2-\frac{1}{2}(S')^2-aS;\,\,0=R_t+\frac{1}{2}
S''+S'R'+aR'
\end{equation}
Changing variables in {\bf (E8)} ($X=(\hbar/\sqrt{m})x$ and $T=\hbar t$) one arrives at
${\bf (E9)}\,\,i\hbar\psi_T=-(\hbar^2/2m)\psi_{XX}-iA\psi_X+V\psi$ where $A=a\hbar/\sqrt{m}$ 
and 
\bq\label{70}
i\hbar R_T+(\hbar^2/m^2)R_XS_X+(\hbar^2/2m^2)S_{XX}+AR_X=0;
\end{equation} 
$$V=-i\hbar S_T+(\hbar^2/2m)R_{XX}+(\hbar^2/2m^2)R_X^2-(\hbar^2/2m^2)S_X^2-AS_X$$
The diffusion equations then take the form
\bq\label{71}
\hbar\phi_T+\frac{\hbar^2}{2m}\phi_{XX}+A\phi_X+\tl{c}\phi=0;\,\,-\hbar\hat{\phi}_T+
\frac{\hbar^2}{2m}\hat{\phi}_{XX}-A\hat{\phi}_X+\tl{c}\hat{\phi}=0;
\end{equation}
$$\tl{c}=-\tl{V}(X,T)-2\hbar S_T-\frac{\hbar^2}{m}S_X^2-2AS_X$$
It is now possible to introduce a role for the quantum potential in this theory.  Thus
from $\psi=exp(R+iS)$ (with $\hbar=m=1$ say) we have $\psi=\rho^{1/2}exp(iS)$ with
$\rho^{1/2}=exp(R)$ or $R=(1/2)log(\rho)$.  Hence $(1/2)(\rho'/\rho)=R'$ and $R''=(1/2)
[(\rho''/\rho)-(\rho'/\rho)^2]$ while the quantum potential is
$Q=(1/2)(\pp^2\rho^{1/2}/\rho^{1/2})=-(1/8)[(2\rho''/\rho)-(\rho'/\rho)^2]$ (cf. 
\eqref{2.4}).  Equation \eqref{68} becomes then
\bq\label{691}
V=-S_t+\frac{1}{8}\left(\frac{2\rho''}{\rho}-\frac{(\rho')^2}{\rho^2}\right)-\frac{1}{2}
(S')^2-aS
\equiv S_t+\frac{1}{2}(S')^2+V+Q+aS=0;
\end{equation}
$$\rho_t+\rho S''+S'\rho'+a\rho'=0\equiv \rho_t+(\rho S')'+a\rho'=0$$
Thus $-2S_t-(S')^2=2V+2Q+2AS$ and one has
\begin{proposition}
The creation-annihilation term $c$ in the diffusion equations (cf. Theorem 3.1) becomes
\bq\label{692}
c=-V-2S_t-(S')^2-2aS'=V+2Q+2a(S-S')
\end{equation}
where Q is the quantum potential.
\end{proposition}
\item
Regarding scale relativity one writes (cf. \eqref{315}
\bq\label{611}
\frac{d_{\pm}}{dt}y(t)=lim_{\gD t\to 0_{\pm}}\left<\frac{\pm y(t\pm\gD t)\mp y(t)}{\gD t}
\right>
\end{equation}
and we collect equations in ($\rho=|\psi|^2$)
\bq\label{612}
dx=b_{+}dt+d\xi_{+}=b_{-}dt+d\xi_{-};<d\xi_{+}^2>=2{\mc D}dt=-<d\xi_{-}^2>
\end{equation}
\bq\label{613}
\frac{d_{+}f}{dt}=(\pp_t+b_{+}\pp+{\mc
D}\pp^2)f;\,\,\frac{d_{-}f}{dt}=(\pp_t+b_{-}\pp-{\mc D}\pp^2)f
\end{equation}
\bq\label{614}
V=\frac{1}{2}(b_{+}+b_{-});\,\,U=\frac{1}{2}(b_{+}-b_{-});\,\,\rho_t+\pp(\rho
V)=0;\,\,U={\mc D}\pp(log(\rho));
\end{equation}
$${\mc V}=V-iU;\,\,d_{{\mc V}}=\frac{1}{2}(d_{+}+d_{-});\,\,d_{{\mc
U}}=\frac{1}{2}(d_{+}-d_{-})$$
\bq\label{615}
\frac{d_{{\mc V}}}{dt}=\pp_t+V\pp;\,\,\frac{d_{{\mc U}}}{dt}={\mc D}\pp^2+U\pp;\,\,\frac
{d'}{dt}=(\pp_t-i{\mc D}\pp^2)+V\pp
\end{equation}
\bq\label{616}
V=2{\mc D}\pp S;\,\,{\mc S}=log(\rho^{1/2})+iS;\,\,\psi=\sqrt{\rho}e^{iS}=e^{i{\mc
S}};\,\,{\mc V}=-2i{\mc D}\pp log(\psi)
\end{equation}
For Lagrangian ${\mc L}=(1/2)m{\mc V}^2-m{\mf U}$ one gets a SE
\bq\label{617}
i\hbar\psi_t=-\frac{\hbar^2}{2m}\pp^2\psi+{\mf U}\psi
\end{equation}
coming from Newton's law ${\bf (E10)}\,\,-\pp{\mf U}=-2i{\mc D}m(d'/dt)\pp
log(\psi)=m(d'/dt){\mc V}$.
\item
The development in Section 3 based on \cite{c29} involves thinking of nonlinear QM as a
fractal Brownian motion with complex diffusion coefficient.  We note {\bf (E10)} corresponds
to {\bf (B58)} and {\bf (B55)} arises in \eqref{615}.  These give rise to 
\bq\label{618}
-\na U=-2im[D\pp_t\na log(\psi)]-2D\na\left(D\frac{\na^2\psi}{\psi}\right)
\end{equation}
Thus putting in a complex diffusion coefficient leads to the NLSE
\bq\label{619}
i\hbar\pp_t\psi=-\frac{\hbar^2}{2m}\frac{\ga}{\hbar}\na^2\psi+U\psi-i\frac{\hbar^2}{2m}
\frac{\gb}{\hbar}(\na log(\psi))^2\psi
\end{equation}
with $\hbar=\ga+i\gb=2mD$ complex.
\item
In \cite{a16} one writes again $\psi=Rexp(iS/\hbar)$ with field equations in the
hydrodynamical picture
\bq\label{620}
d_t(m_0\rho v)=\pp_t(m_0\rho v)+\na(m_0\rho v)=-\rho\na(u+Q);\,\,\pp_t\rho+\na\cdot(\rho v)=0
\end{equation}
where $Q=-(\hbar^2/2m_0)(\gD\sqrt{\rho}/\sqrt{\rho})$.  One works with the Nottale approach
as above with $d_v\sim d_{{\mc V}}$ and $d_u\sim d_{{\mc U}}$ (cf. \eqref{615}).  One assumes
that the velocity field from the hydrodynamical model agrees with the real part $v$ of the
complex velocity $V=v-iu$ so (cf. \eqref{614}) $v=(1/m_0)\na s\sim 2D\pp s$ and $u=-(1/m_0)
\na\gs\sim D\pp log(\rho)$ where $D=\hbar/2m_0$.  In this context the quantum potential
$Q=-(\hbar^2/2m_0)\gD\sqrt{\rho}/\sqrt{\rho}$ becomes ${\bf (E11)}\,\,Q=-m_0D\na
u-(1/2)m_0u^2\sim -(\hbar/2)\pp u-(1/2)m_0u^2$.  Consequently Q arises from the fractal
derivative and the nondifferentiability of spacetime.  Further one can relate $u$ (and hence
Q) to an internal stress tensor {\bf (D25)} whereas the $v$ equations correspond to systems
of Navier-Stokes type.  Note here that \eqref{59} involves a term relating the stress
tensor $\Pi$ and Q directly.
\end{enumerate}

\section{CONCLUSIONS}
\renewcommand{\theequation}{7.\arabic{equation}}
\setcounter{equation}{0}

One feature either exhibited or suggested in the examples displayed involves the role of a
quantum potential in either quantization or ``classicalization" of certain systems of
equations of hydrodynamic type.  Now with numbers referring to Section 6 we have:
\begin{enumerate}
\item
One arrived at an equation of hydrodynamic type directly from the SE upon addition of a
pressure term which served to augment the original potential V (however this could have
simply been included in V).  On the other hand Q does not appear in the SE but is generated
by the decomposition $\psi=Rexp(iS/\hbar)$
\item
In a general statistical mechanical approach, with the dynamics determined by classical
kinetics in a phase space, the quantum potential has an interpretation in terms of an
internal stress tensor for a quantum fluid.  The equations are again described in terms
of a probability density $\rho$ and a phase factor $S$.
\item
In \#3-\#4 one takes a classical statistical ensemble with S a momentum potential and
expresses momentum fluctuations in terms of Fisher information; this leads to a SE with
quantization term Q expressed as Fisher information.  In Remarks 2.6-2.8 we show how Fisher
information, entropy, and the quantum potential are mutually entangled (cf. also
\cite{d99}).   In
\eqref{5.59}-\eqref{5.60} (based on \cite{g10}) we see how the Euler equation
$(\pp_t+v\cdot\na)v=(F/m)-(\gD P/\rho)$ (where P is a pressure term) is related to the
quantized form $(\pp_t+v\cdot\na)v= (F/m)-\na Q$ arising from a SE.
\item
The Nagasawa-Nelson approach in \#5 views matters rather differently in showing the
equivalence of the SE to a pair of diffusion equations.  The full theory is very elegant and
extends to singular situations, etc. (cf. \cite{n9,n15}).  It would be of interest here to
further examine the quantum potential in this context.
\item
In \#6-\#8 one arrives at a pair of equations by virtue of the ``fractal" structure of
space (where fractal here simply means that nondifferentiable paths are considered which
generate a complex velocity).  In \cite{a16} (as exhibited in \#8) one relates the quantum
potential to the velocity $u$, showing its origin in the ``fractal" derivative idea.
\end{enumerate}
\indent
We emphasize that in fact the quantum potential comes up in a serious manner in the Bohm
theory, with refinements as in \cite{ch,c2,c3,c4,c5,c6,d10,d4,d5,f2,f3,f7,f8,h99,h98}.
In fact, given that trajectories are at the base of this theory one can forsee a fractal
Bohm theory in the future (cf. \cite{h95,n57}).  On the other hand one can make convincing
arguments for fields as the fundamental objects (except perhaps in the Bohmian type theories)
with particules ``emerging" (cf. \cite{h97,w12}) as in quantum field theory (or perhaps
via ripples or fractal structure in spacetime itself).
\\[3mm]\indent
It is not entirely clear how to handle derivatives in statistical or fractal theories. 
There are of course many powerful techniques available for Brownian motion and stochastic
differential equations and there is a developing literature about differential calculus on
fractals.  Random walks and general discretization methods are also useful.  Somehow one
would like to imagine that the formal power of calculus (and duality via distribution
like theories) might be strong enough to override the microscopic details about the domains
of differential operators.  Perhaps the coordinate derivative operators in situations 
such as \#6-\#8 could be defined so that their domains are various fractal sets densely
embedded in ${\bf R}^n$ (in this connection see e.g. \cite{c37,k99,k12,m91,n59,o8,p12,s91}). 
In the end the most attractive formulation would seem to be some (more or less rigorous)
version of a Feynmann path integral where precise definitions of the path space are not
critical.

\end{document}